\documentclass[twocolumn]{aastex63}
\usepackage[normalem]{ulem}
\submitjournal{ApJ}

\shorttitle{Gaia 20eae}
\shortauthors{Ghosh et al.}
\graphicspath{{./}{figures/}}
\begin{document}

\title{Gaia 20eae: A  newly discovered episodically accreting young star}

        \author{Arpan Ghosh}
        \affil{Aryabhatta Research Institute of Observational Sciences (ARIES),
        Manora Peak, Nainital 263 001, India}
        \affil{School of Studies in Physics and Astrophysics, Pandit Ravishankar Shukla University, Raipur 492010, Chhattisgarh, India},
        \author{Saurabh Sharma}
        \affil{Aryabhatta Research Institute of Observational Sciences (ARIES),
        Manora Peak, Nainital 263 001, India},
        \author{Joe P. Ninan}
        \affil{Department of Astronomy and Astrophysics, The Pennsylvania State University, 525 Davey Laboratory, University Park, PA
16802, USA}
        \affil{Center for Exoplanets and Habitable Worlds, The Pennsylvania State University, 525 Davey Laboratory, University Park, PA
16802, USA}
        \author{Devendra K. Ojha}
        \affil{ Department of Astronomy and Astrophysics, Tata Institute of Fundamental Research (TIFR), Mumbai 400005, Maharashtra, India}
        \author{Bhuwan C. Bhatt}
        \affil{Indian Institute of Astrophys., II Block, Koramangala, Bangalore 560 034, India}
        \author{Shubham Kanodia}
        \affil{Department of Astronomy and Astrophysics, The Pennsylvania State University, 525 Davey Laboratory, University Park, PA
16802, USA}
       \affil{Center for Exoplanets and Habitable Worlds, The Pennsylvania State University, 525 Davey Laboratory, University Park, PA
16802, USA}
       \author{Suvrath Mahadevan}
       \affil{Department of Astronomy and Astrophysics, The Pennsylvania State University, 525 Davey Laboratory, University Park, PA
16802, USA}
       \affil{Center for Exoplanets and Habitable Worlds, The Pennsylvania State University, 525 Davey Laboratory, University Park, PA
16802, USA}
       \author{Gudmundur Stefansson}
       \affil{Princeton University, Princeton, United States}
       \author{R. K. Yadav}
       \affil{National Astronomical Research Institute of Thailand, Chiang Mai, 50200, Thailand}
       \author{A. S. Gour}
       \affil{School of Studies in Physics and Astrophysics, Pandit Ravishankar Shukla University, Raipur 492010, Chhattisgarh, India}
       \author{Rakesh Pandey}
       \affil{Aryabhatta Research Institute of Observational Sciences (ARIES),
        Manora Peak, Nainital 263 001, India}
       \affil{School of Studies in Physics and Astrophysics, Pandit Ravishankar Shukla University, Raipur 492010, Chhattisgarh, India}
       \author{Tirthendu Sinha}
       \affil{Aryabhatta Research Institute of Observational Sciences (ARIES),
        Manora Peak, Nainital 263 001, India}
        \affil{Kumaun University, Nainital 263001, India}
       \author{Neelam Panwar}
       \affil{Aryabhatta Research Institute of Observational Sciences (ARIES),
        Manora Peak, Nainital 263 001, India}
       \author{John P. Wisniewski}
       \affil{Homer L. Dodge Department of Physics and Astronomy, University of Oklahoma, Norman, OK 73019, USA}
       \author{Caleb I. Ca\~nas}
       \affil{Department of Astronomy and Astrophysics, The Pennsylvania State University, 525 Davey Laboratory, University Park, PA
16802, USA}
       \affil{Center for Exoplanets and Habitable Worlds, The Pennsylvania State University, 525 Davey Laboratory, University Park, PA
16802, USA}
       \affiliation{NASA Earth and Space Science Fellow}
       \author{Andrea S. J. Lin}
       \affil{Department of Astronomy and Astrophysics, The Pennsylvania State University, 525 Davey Laboratory, University Park, PA
16802, USA}
       \affil{Center for Exoplanets and Habitable Worlds, The Pennsylvania State University, 525 Davey Laboratory, University Park, PA
16802, USA}
      \author{Arpita Roy}
      \affil{Space Telescope Science Institute, 3700 San Martin Drive, Baltimore, MD 21218, USA}
      \affil{Department of Physics and Astronomy, Johns Hopkins University, 3400 North Charles Street, Baltimore, MD 21218, USA}
      \author{Fred Hearty}
      \affil{Department of Astronomy and Astrophysics, The Pennsylvania State University, 525 Davey Laboratory, University Park, PA
16802, USA}
      \affil{Center for Exoplanets and Habitable Worlds, The Pennsylvania State University, 525 Davey Laboratory, University Park, PA
16802, USA}
      \author{Lawrence Ramsey}
      \affil{Department of Astronomy and Astrophysics, The Pennsylvania State University, 525 Davey Laboratory, University Park, PA
16802, USA}
      \affil{Center for Exoplanets and Habitable Worlds, The Pennsylvania State University, 525 Davey Laboratory, University Park, PA
16802, USA}
      \author{Paul Robertson}
      \affil{Department of Physics and Astronomy, University of California Irvine, Irvine, CA 92697, USA}
      \author{Christian Schwab}
      \affil{Department of Physics and Astronomy, Macquarie University, Balaclava Road, North Ryde, NSW 2109, Australia}



\begin{abstract}
The Gaia Alert System issued an alert on 2020 August 28, on Gaia 20eae when its light curve showed a $\sim$4.25 magnitude outburst. 
We present multi-wavelength photometric and spectroscopic follow-up observations of this source since 2020 August and  
identify it as the newest member of the FUor/EXor family of sources. 
We find that the present brightening of Gaia 20eae is not due to the dust clearing event 
but due to an intrinsic change in the spectral energy distribution. 
The light curve of Gaia 20eae shows a transition stage during which most of its brightness ($\sim$3.4 mag) has occurred 
at a short timescale of 34 days with a rise-rate of 3 mag/month.
Gaia 20eae has now started to decay at a rate of 0.3 mag/month.
We have detected a strong P Cygni profile in H$\alpha$ which indicates the presence of winds originating from regions close to the accretion.
We find signatures of very strong and turbulent outflow and accretion in Gaia 20eae during this outburst phase.
We have also  detected a red-shifted absorption component in all the Ca II IR triplet lines consistent with signature of hot in-falling gas in the magnetospheric accretion funnel. This enables us to constrain the viewing angle with respect to the accretion funnel.
Our investigation of Gaia 20eae points towards magnetospheric accretion being the phenomenon for the current outburst.

\end{abstract}


%% to the KEY in the \bibitem in the reference list below. https://www.overleaf.com/project/5f68aa3add3f7200013222aa

\section{Introduction}

Episodic accretion onto low-mass pre-main sequence (PMS) stars is no longer considered an oddity. 
It is now considered as one of the important stages in the grand scheme of evolution of the low-mass PMS
 stars, even though it is a poorly understood phenomenon. The short outburst timescales compared to the millions of years spent in the 
formation stage of these PMS stars makes these events extremely rare, although statistically each PMS star is expected to 
experience $\sim$50 such short duration outbursts during its formation stages \citep{2013MNRAS.430.2910S}. 
The outburst durations, although short in timescales, are capable of delivering a substantial fraction of circumstellar mass onto
the central PMS star \citep{2006ApJ...650..956V}. These events have been observed to span the entire age range of young stars starting from 
the embedded Class 0 sources to the Class\,{\sc ii} sources \citep{2015ApJ...800L...5S}. 
Based on the outburst timescales and spectroscopic features, these classes of sources have been classically 
divided into two categories: FUors which experience a luminosity outburst of 4-5 mag that last for several decades and containing only
absorption lines in their spectra and EXors experiencing a luminosity outburst of 2-3 mag which last for a timescale of few months to few 
years and 
contains emission lines in their spectra \citep{1977ApJ...217..693H,1996ARA&A..34..207H, 1998apsf.book.....H}. 
The physical origin of the sudden enhancement of accretion rate is not yet clear. However,  a variety of models ranging from thermal instability, magneto-rotational instability, combination of magneto-rotational instability and gravitational instability, disc fragmentation to external perturbations have been proposed \citep{2014prpl.conf..387A}. To arrive at a general consensus about the 
physics behind such sudden enhancement of accretion rates, a large sample of FUor/EXor sources are required to test the above instability models. 
However, only about 25 FUor/EXor sources have been discovered so far \citep[][]{2014prpl.conf..387A}.
Therefore, any newly discovered source provides an important test-bed to probe the various physical aspects of episodic accretion and their comparison with the previous sources.

   The $Gaia$ Photometric Alert System \citep{2012gfss.conf...21W,2013RSPTA.37120239H} is dedicated towards issuing transient alerts. 
   Previously, it has issued three proven alerts for the eruptive young stars: 
   Gaia 17bpi \citep{2018ApJ...869..146H}, Gaia 19ajj \citep{2019AJ....158..240H} and Gaia 18dvy \citep{2020ApJ...899..130S}.
   Among these, Gaia 17bpi and Gaia 18dvy are classified as FUors while Gaia 19ajj has been classified as an EXor, with its spectral features
   similar to that of V2492 Cyg \citep{2019AJ....158..240H}. 
   The $Gaia$ alert system issued an notification on 2020 August 28 about Gaia 20eae with a transient identification number of  AT2020nrs stating that it has undergone a 4.6 magnitude outburst. 
	The rise timescales and the amplitude of the outburst suggest that this should be an FUor/EXor phenomenon. 
	We have carried out optical and near-infrared (NIR) photometric and spectroscopic observations and combined them with the archival 
	optical and infrared (IR) data to identify the outburst features of Gaia 20eae.  In this paper, we present the initial findings of this source.
	Section \ref{pt0} provides details on the location and distance of Gaia 20eae.
	Section \ref{pt1} describes about the observations and the data reduction procedures in detail. In
	Section \ref{pt2} we describe the results that have been obtained while in Section \ref{pt3} we conclude by our understanding of the present outburst in the context of the FUor/EXor phenomena.

\begin{figure*}[h]
\centering
\includegraphics[width=0.65\textwidth]{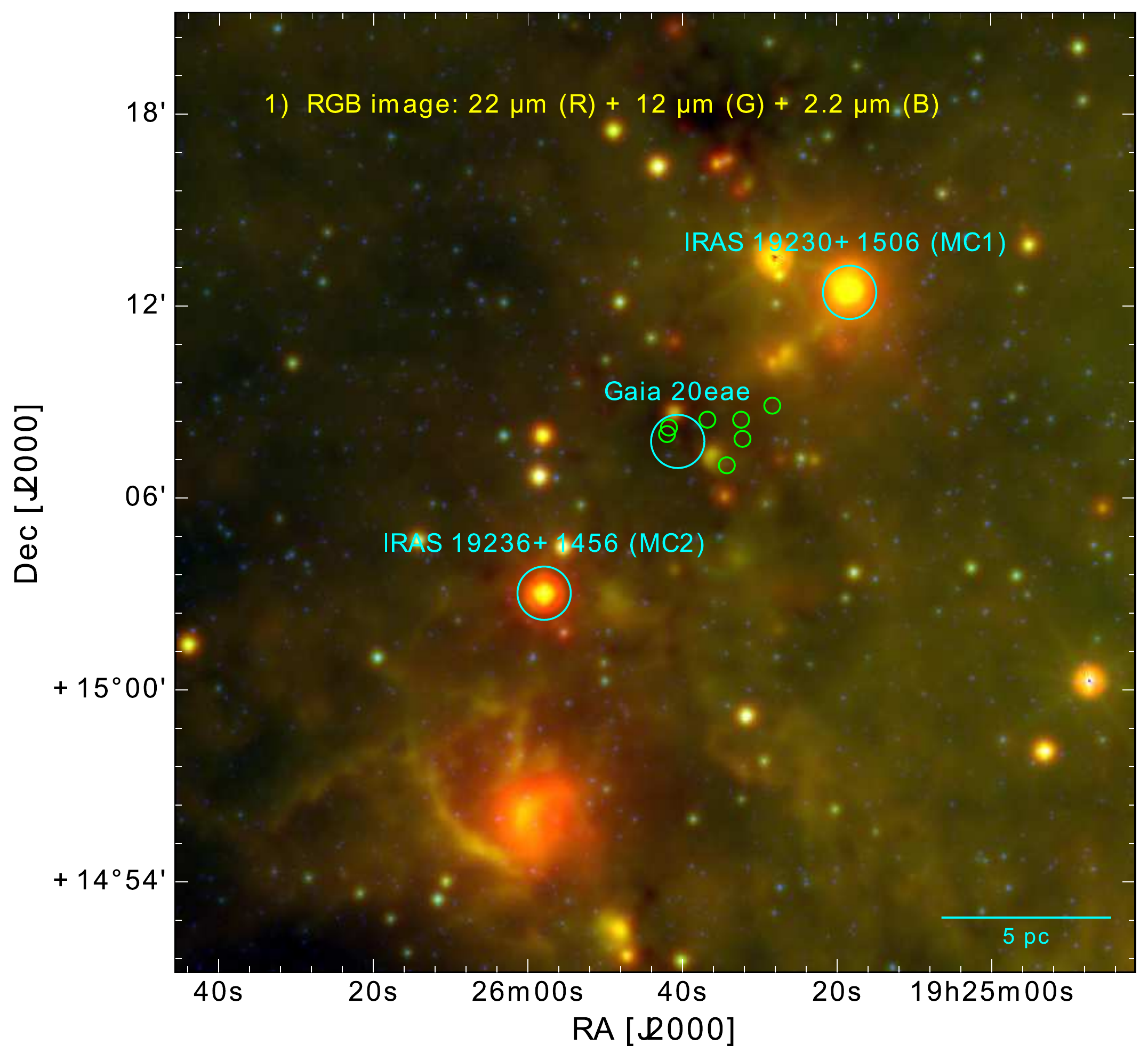}
\includegraphics[width=0.31\textwidth]{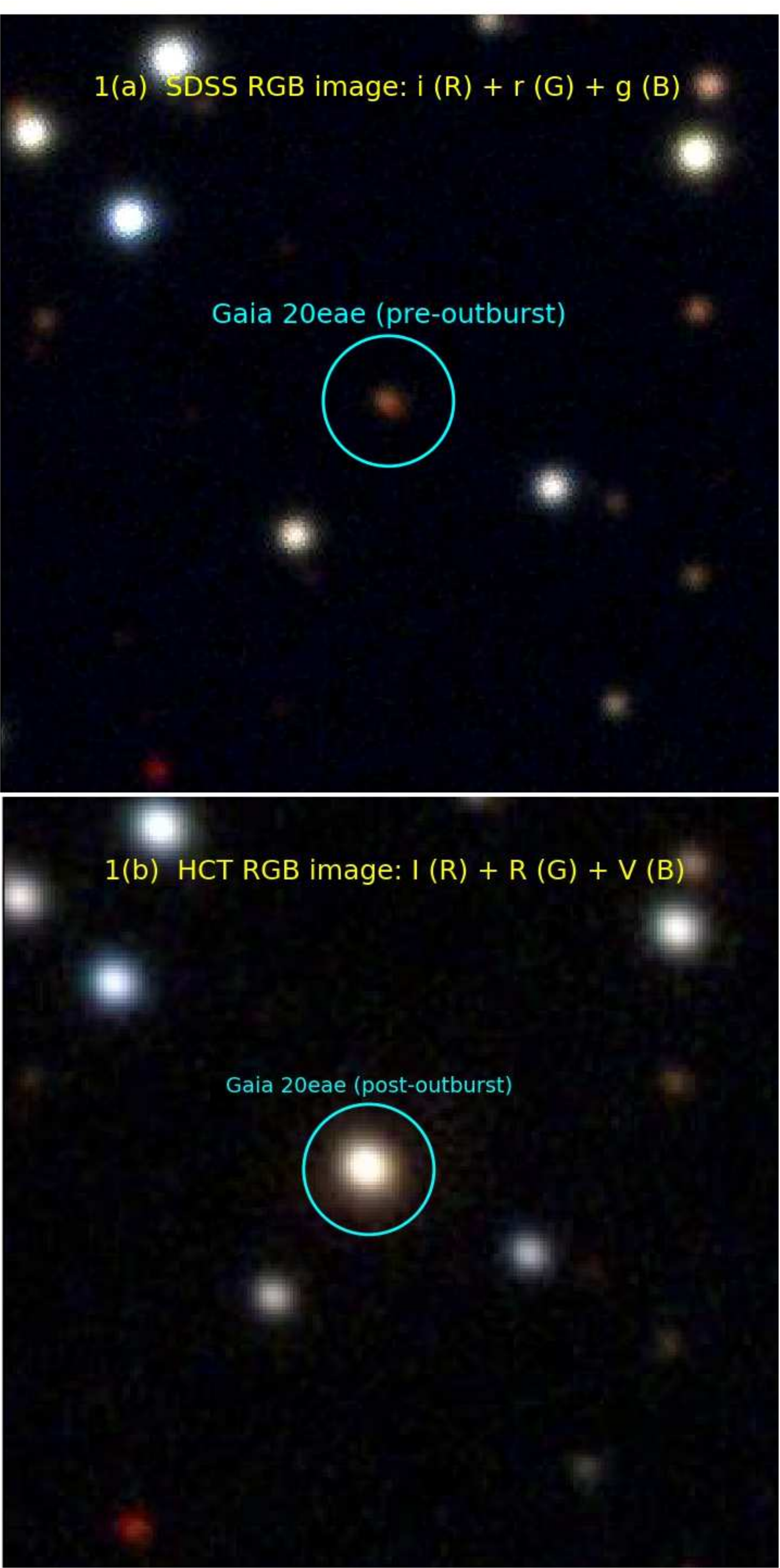}
\caption{\label{fima} Color-composite image obtained by using the WISE 22 $\mu$m (red),  WISE 12 $\mu$m (green) and 2MASS 2.2 $\mu$m (blue)
images of the $\sim30^\prime\times 30^\prime$ Field of View (FOV) around Gaia 20eae.
The locations of Gaia 20eae, IRAS 19230+1506 and IRAS 19236+1456 \citep{2017ApJ...839..113R} are shown by cyan circles. 
Locations of standard stars from the ZTF sky survey  is also shown with green circles.
Sub-panels 1(a) and 1(b) show the pre-outburst and post-outburst phases  of Gaia 20eae
in optical color composite image taken from SDSS and HCT, respectively.}
\end{figure*}

\section{G\lowercase{aia} 20\lowercase{eae}: Location and distance} \label{pt0}

The Gaia 20eae   ($\alpha$$_{2000}$ =19$^{h}$25$^{m}$40$^{s}.61$,
$\delta$$_{2000}$ = +15$^\circ$07$^\prime$46$^{\prime\prime}.5$; $l = 50.258492^\circ, b= -00.507730^\circ$)
is located near the edge of the W51 star-forming complex. The W51 star-forming region is known to be one of the most massive and active star-forming
sites of our Galaxy located at about 5 kpc distance from us \citep{2017arXiv170206627G}.
\citet{2017ApJ...839..113R} have listed two molecular cloud MC1 (size$\sim15^\prime \times15^\prime$) and MC2 (size$\sim30^\prime \times36^\prime$)
in this direction at two different distances of $1.3\pm0.2$ kpc and $3.4\pm0.4$ kpc, respectively.
The distance of these molecular clouds are derived kinematically using the CS ($2\rightarrow1$) line velocities \citep{2004A&A...426...97F}.
The MC1 and MC2 are also associated with IRAS sources IRAS 19230+1506 and IRAS 19236+1456, respectively \citep{2017ApJ...839..113R}.
These molecular clouds do not have any optical nebula associated with them, indicating that the star-formation activity has started recently.
Also, the IRAS sources have ultra-compact H\,{\sc ii} region (UCH{\sc ii}) colors indicating that these molecular clouds are high mass star-forming regions.
In Figure \ref{fima}, we show the  location of  Gaia 20eae along with the IRAS sources in the IR color-composite image generated from 
 WISE 22 $\mu$m (red),  WISE 12 $\mu$m (green), and 2MASS 2.2 $\mu$m (blue) images.
Heated dust grains (22 $\micron$ emission) can be seen at several places including at the IRAS locations of sources.
The warm dust towards the south of Gaia 20eae is surrounded by 12 $\mu$m emission which covers the prominent
Polycyclic Aromatic Hydrocarbons (PAH) features at 11.3 $\mu$m, indicative of photon dominant region (PDR)
under the influence of feedback from massive stars \citep[see e.g.][]{2004ApJ...613..986P}.
This indicates that the Gaia 20eae is located at a site showing signatures of recent star-formation activities.
Until the release of data from the Gaia mission\footnote{https://sci.esa.int/web/gaia}, there was no direct measurement of distance of the Gaia 20eae. Recently, adding corrections to the Gaia data release 3 (DR3) parallax by using
the Bayesian inference approach to account for the
non-linearity of the transformation and the asymmetry of the resulting probability distribution,
\citet{2021AJ....161..147B} estimated the distances of stars in our Galaxy. 
Therefore, for Gaia 20eae, we have adopted a distance of 3.2 $\pm$ 1 kpc as estimated by \citet{2021AJ....161..147B}. 
Since, molecular cloud MC2 is located in the same direction at a distance of $3.4\pm0.4$ kpc \citep{2017ApJ...839..113R}, 
Gaia 20eae's seems to be associated with MC2.
This makes it the farthest discovered FUor/EXor type source till date.
Almost all the previously discovered FUors/EXors are located at a distance of $\sim$1 kpc or less.

\begin{table*}
\centering
\caption{Log of Photometric and Spectroscopic Observations.}
\label{tab:obs_log}
\begin{tabular}{@{}rrrrr@{}}
\hline 
Telescope/Instrument & Date  & Julian Day & Filters/Grisms & Exposure(sec) $\times$ Number of frames  \\
\hline
2.0m HCT HFOSC  & 2020 Aug 29 & 2459091 & $Gr 7, Gr 8$ & 2100$\times$1, 1800$\times$1  \\
2.0m HCT HFOSC  & 2020 Aug 30 & 2459092 & $Gr 7$ &  2400$\times$1\\
2.0m HCT HFOSC  & 2020 Aug 31 & 2459093 & $B, V, R, I, Gr 7$ & 120$\times$1,60$\times$1,30$\times$1,30$\times$2, 2400$\times$1\\
2.0m HCT HFOSC  & 2020 Sep 01 & 2459094 & $Gr 7, Gr8$ & 2400$\times$1, 2400$\times$1 \\
2.0m HCT HFOSC  & 2020 Sep 07 & 2459100 & $B, V, R, I, Gr7,  Gr8$ & 120$\times$2, 60$\times$2, 30$\times$2, 30$\times$2, 2400$\times$1, 1800$\times$1\\
10.0m HET HPF   & 2020 Sep 11 & 2459104 & NIR Cross dispersed echelle & 617.7$\times$2\\
2.0m HCT HFOSC  & 2020 Sep 12 & 2459105 & $B, V, R, I$ & 120$\times$2, 60$\times$2, 30$\times$2, 30$\times$2 \\
10.0m HET LRS2  & 2020 Sep 12 & 2459105 & Optical low resolution spectra & 157.7$\times$2\\
2.0m HCT HFOSC  & 2020 Sep 14 & 2459107 & $B, V, R, I, Gr7,  Gr8$ & 180$\times$4, 60$\times$4, 30$\times$3, 30$\times$2,2700$\times$1, 2700$\times$1\\
1.3m DFOT 2KCCD & 2020 Oct 11 & 2459134 & $V, R, I$ & 10$\times$30, 10$\times$12, 10$\times$12\\
1.3m DFOT 2KCCD & 2020 Oct 12 & 2459135 & $V, R, I$ & 10$\times$20, 10$\times$8, 10$\times$8\\
1.3m DFOT 2KCCD & 2020 Oct 13 & 2459136 & $V, R, I$ & 15$\times$20, 15$\times$8, 15$\times$8\\
0.5m ARCSAT     & 2020 Oct 12 & 2459135 & $g,r, i, z$ & 300$\times$1,120$\times$6,120$\times$6,180$\times$6\\ 
0.5m ARCSAT     & 2020 Oct 13 & 2459136 & $g,r, i, z$ & 360$\times$5,300$\times$5,240$\times$11,240$\times$5 \\
0.5m ARCSAT     & 2020 Oct 18 & 2459141 & $g,r, i, z$ & 900$\times$2,300$\times$3,240$\times3$\\
1.3m DFOT 2KCCD & 2020 Oct 19 & 2459142 & $B, V, R, I$ & 60$\times$4, 60$\times$3, 60$\times$1, 60$\times$1\\ 
1.3m DFOT 2KCCD & 2020 Oct 20 & 2459143 & $B, V, R, I$ & 60$\times$4, 60$\times$3, 60$\times$1, 60$\times$1\\  
3.6m DOT  TANSPEC & 2020 Oct 21 & 2459144 &$J,H, K$    & 10$\times$4$\times$ 5 dither \\  
3.6m DOT  TANSPEC & 2020 Oct 24 & 2459147 & Cross Dispersed spectra & 150$\times$8 \\
3.6m DOT  TANSPEC & 2020 Nov 06 & 2459160 & Cross Dispersed spectra & 150$\times$8 \\
1.3m DFOT 2KCCD & 2020 Nov 08 & 2459162 & $B, V, R, I$ & 240$\times$2, 150$\times$1, 60$\times$1, 60$\times$1 \\  
1.3m DFOT 2KCCD & 2020 Nov 13 & 2459167 & $B, V, R, I$ & 240$\times$7, 150$\times$7, 60$\times$7, 60$\times$7\\  
1.3m DFOT 2KCCD & 2020 Nov 14 & 2459168 & $B, V, R, I$ & 240$\times$4, 150$\times$2, 60$\times$1, 60$\times$1\\  
1.3m DFOT 2KCCD & 2020 Dec 07 & 2459191 & $V, R, I$ &  60$\times$3,60$\times$3,60$\times$3\\   
\hline
\end{tabular}
\end{table*}

\section{Observation and Data Reduction} \label{pt1} 

\subsection{Photometric data}

\subsubsection{Present data}

We have monitored Gaia 20eae photometrically in optical bands at 16 
different epochs with the Himalayan Faint Optical Spectrograph Camera\footnote{\url{https://www.iiap.res.in/iao/hfosc\_details.html}} (HFOSC, 4 epochs) 
on 2m Himalayan \textit{Chandra} Telescope (HCT), Hanle, India, ANDOR 2K CCD (9 epochs) 
on 1.3m Devasthal Fast Optical Telescope (DFOT), Nainital India,  
and FlareCam 1K CCD\footnote{\url{https://www.apo.nmsu.edu/Telescopes/ARCSAT/Instruments/arcsat_instruments.html}} on 0.5m ARC Small Aperture Telescope (ARCSAT, 3 epochs), New Mexico\footnote{Apache Point Observatory (APO) located in Sunspot, New Mexico which is operated by the Astrophysical Research Consortium (ARC). ARCSAT is a 0.5m Classical Cassegranian Telescope, formerly known as the SDSS Photometric Telescope (PT).} from 2020 August to December.
We have also obtained the near infrared (NIR) photometric data of Gaia 20eae during its 
outburst state using the TIFR-ARIES Near Infrared Spectrometer \citep[TANSPEC;][]{2018BSRSL..87...58O} mounted 
on the 3.6m Devasthal Optical Telescope (DOT), Nainital, India on the night of 2020 October 24. 
Table \ref{tab:obs_log} provides the complete log of photometric observations presented in  this work.

We have used standard data reduction procedures for the image cleaning, photometry, and astrometry \citep[for details, see ][]{2020MNRAS.498.2309S}.
We have derived following color transformation equations using the
available magnitudes in different filters (i.e., APASS DR10 archive\footnote{https://www.aavso.org/download-apass-data} or
Two Micron All Sky Survey (2MASS) archive\footnote{https://irsa.ipac.caltech.edu/Missions/2mass.html}) of all the stars in the frame
at epoch JD $=$ 2459141 (for optical) and JD $=$ 2459144 (for NIR).

\begin{equation} \label{eqn1}
B-V = 1.06\pm0.02 \times (b-v) - 0.71\pm0.03 
\end{equation}

\begin{equation} \label{eqn2}
V-R = 0.77\pm0.02 \times (v-r) - 0.21\pm0.02
\end{equation}

\begin{equation} \label{eqn3}
R-I = 0.88\pm0.04 \times (r-i) + 0.66\pm0.02
\end{equation}

\begin{equation} \label{eqn4}
R-r = 0.09\pm0.04 \times (V-R) + 2.35\pm0.02
\end{equation}

\begin{equation} \label{eqn5}
J-H = 0.90\pm0.06 \times (j-h) + 0.05\pm0.09
\end{equation}

\begin{equation} \label{eqn6}
H-K = 1.00\pm0.07 \times (h-k) + 0.52\pm0.05
\end{equation}

\begin{equation} \label{eqn7}
J-j = 0.01\pm0.04 \times (J-H) - 1.20\pm0.04
\end{equation}

In order to calibrate the photometry of Gaia 20eae at other epochs, we have used these equations 
with intercept estimated from a set of 7 non-variable standard stars (see Table \ref{tab:std_log}).
These non-variable standard stars 
were identified from the Zwicky Transient Facility (ZTF) sky survey \citep{Bellm_2018} based on their zr band light curves 
(LCs\footnote{https://irsa.ipac.caltech.edu/Missions/ztf.html}). 
Table 2 lists the magnitudes of Gaia 20eae in different filters in different epochs of our observations.

\begin{table*}
\centering
\tiny
\caption{Photometric magnitudes of Gaia 20eae in different filters using the present observations.}
\label{tab:phot_tab}
\begin{tabular}{@{}rrrrrrrrrrrr@{}}
\hline
JD & $B$     & $V$   & $R_c$   & $I_c$   & $g$  & $r$  & $i$   & $z$   &  $J$  & $H$  & $K_{s}$\\ 
   & (mag)   & (mag) & (mag) & (mag) &(mag) &(mag) & (mag) & (mag) & (mag) & (mag)& (mag) \\
\hline
2459093 & 18.60$\pm$0.03 &  16.49$\pm$0.01 &  15.23$\pm$0.01 & 14.33$\pm$0.01 & $-$           & $-$           & $-$           & $-$             & $-$    & $-$    & $-$\\   
2459100 & 18.88$\pm$0.01 &  16.74$\pm$0.01 &  15.41$\pm$0.01 & 14.45$\pm$0.01 & $-$           & $-$           & $-$           & $-$             & $-$    & $-$    & $-$ \\ 
2459105 & 18.63$\pm$0.01 &  16.52$\pm$0.01 &  15.23$\pm$0.01 & 14.38$\pm$0.01 & $-$           & $-$           & $-$           & $-$             & $-$    & $-$    & $-$  \\
2459107 & 18.16$\pm$0.01 &  16.47$\pm$0.01 &  15.20$\pm$0.01 & 14.27$\pm$0.01 & $-$           & $-$           & $-$           & $-$             & $-$    & $-$    & $-$   \\
2459134 & $-$            &  17.43$\pm$0.01 &  16.11$\pm$0.01 & 14.71$\pm$0.02 & $-$           & $-$           & $-$           & $-$             & $-$    & $-$    & $-$    \\
2459135 & $-$            &  17.19$\pm$0.01 &  15.52$\pm$0.01 & 14.66$\pm$0.02 & 18.01$\pm$0.07& 16.24$\pm$0.01& 15.02$\pm$0.01& 14.22$\pm$0.02  & $-$    & $-$    & $-$     \\ 
2459136 & $-$            &  17.47$\pm$0.02 &  15.68$\pm$0.01 & 14.70$\pm$0.01 & 17.94$\pm$0.03& 16.17$\pm$0.04& 15.03$\pm$0.02& 14.23$\pm$0.02  & $-$    & $-$    & $-$      \\  
2459141 & $-$            &  $-$            &  $-$            & $-$            & 17.86$\pm$0.04& 16.26$\pm$0.07& 15.11$\pm$0.04& $-$             & $-$    & $-$    & $-$      \\    
2459142 & 19.21$\pm$0.03 &  17.38$\pm$0.01 &  16.08$\pm$0.01 & 14.72$\pm$0.01 & $-$           & $-$           & $-$           & $-$             & $-$    & $-$    & $-$       \\ 
2459143 & 19.12$\pm$0.04 &  17.14$\pm$0.01 &  15.91$\pm$0.02 & 14.61$\pm$0.01 & $-$           & $-$           & $-$           & $-$             & $-$    & $-$    & $-$        \\        
2459144 & $-$            &  $-$            &  $-$            & $-$            & $-$           & $-$           & $-$           & $-$             & 12.41$\pm$0.02  & 11.27$\pm$0.03  & 10.40$\pm$0.03        \\
2459162 & 19.54$\pm$0.03 &  17.74$\pm$0.01 &  16.41$\pm$0.01 & 15.08$\pm$0.01 & $-$           & $-$           & $-$           & $-$             & $-$    & $-$    & $-$            \\  
2459167 & 18.67$\pm$0.01 &  16.96$\pm$0.01 &  15.73$\pm$0.01 & 14.45$\pm$0.01 & $-$           & $-$           & $-$           & $-$             & $-$    & $-$    & $-$             \\
2459168 & 18.77$\pm$0.03 &  17.08$\pm$0.01 &  15.83$\pm$0.01 & 14.54$\pm$0.01 & $-$           & $-$           & $-$           & $-$             & $-$    & $-$    & $-$              \\
2459191 & $-$            &  17.52$\pm$0.01 &  16.20$\pm$0.01 & 14.89$\pm$0.01 & $-$           & $-$           & $-$           & $-$             & $-$    & $-$    & $-$               \\
\hline
\end{tabular}
\end{table*}

\begin{table}
\centering
\caption{Coordinates of the local standard stars.}
\label{tab:std_log}
\begin{tabular}{@{}rrrrr@{}}
\hline
ID & $\alpha$$_{2000}$  & $\delta$$_{2000}$ & $zr$$\pm$$\sigma$ \\
      & (degrees)           &  (degrees)         & (mag)    \\
\hline
1 &  291.423929  & +15.136802  & 15.51$\pm$0.01\\
2 &  291.424941  & +15.133386  & 16.01$\pm$0.01\\
3 &  291.403179  & +15.140891  & 15.50$\pm$0.01\\
4 &  291.384237  & +15.130902  & 14.95$\pm$0.01\\
5 &  291.384966  & +15.140891  & 15.02$\pm$0.01\\
6 &  291.392691  & +15.117122  & 12.75$\pm$0.01\\
7 &  291.368150  & +15.148200  & 13.10$\pm$0.01\\
\hline
\end{tabular}
\end{table}

\subsubsection{Archival data}

We have also obtained the photometric data from the time domain Gaia sky survey  
\citep{2016A&A...595A...1G,2018A&A...616A...1G}.
Gaia sky survey maps the sky in $G$ band to look out for the transients and regularly updates on 
their Gaia Alert Index website\footnote{http://gsaweb.ast.cam.ac.uk/alerts/alertsindex}. 
We have obtained the $G$ band photometric data provided by the Gaia survey
at its data archive\footnote{https://gea.esac.esa.int/archive/}.
 
We have also acquired the pre-outburst $g,r,i,z$ band archival data from the Panoramic Survey Telescope and Rapid Response System (Pan-STARRS; PS1). 
The details about the PS1 surveys and latest data products are given in \citet{2016arXiv161205560C}.
We have downloaded the point source catalog from  the data release 2 of the PS1\footnote{http://catalogs.mast.stsci.edu/}.
  
 Gaia 20eae was observed by the Zwicky ZTF sky survey \citep{Bellm_2018}. We obtained the archival $zr$ band photometric data of ZTF available from  the 
NASA/IPAC Infrared Science Archive\footnote{https://www.ipac.caltech.edu}. We also obtained the recent 
photometric data in $zg$ and $zr$ bands of the observation made by the ZTF survey from 
Lasair 2.0\footnote{https://lasair.roe.ac.uk/}, a community broker service to access, visualize and extract science data.

 We obtained the pre-outburst mid-infrared (MIR) magnitudes of Gaia 20eae from the $Spitzer$ 
archive\footnote{https://irsa.ipac.caltech.edu/Missions/spitzer.html} 
in 3.6 $\mu$m, 4.5 $\mu$m, 5.8 $\mu$m, 8.0 $\mu$m and
24.0 $\mu$m wave bands. Pre-outburst magnitudes of Gaia 20eae are also obtained from the $WISE$ 
archive\footnote{https://irsa.ipac.caltech.edu/Missions/wise.html} in 3.4 $\mu$m, 4.6 $\mu$m, 12 $\mu$m and
22 $\mu$m wave bands.  Outburst magnitudes of Gaia 20eae are obtained from the $WISE/NEOWISE$ 
survey\footnote{https://irsa.ipac.caltech.edu/Missions/wise.html}
 in 3.4 $\mu$m and 4.6 $\mu$m wave bands. 

\subsection{Spectroscopic data} 

\subsubsection{Medium resolution Optical/NIR Spectroscopy}

A photometric alert was issued by the Gaia alert system named AT2020nrs on 2020 August 28, 1:28 p.m. UTC. We immediately followed it using  a medium 
resolution (R$\sim$2000) spectrograph `HFOSC' mounted on the 2m HCT starting from 2020 August 29 itself.
Using the $Gr 7$ and $Gr 8$ grisms
of HFOSC, our spectroscopic observations spanned the optical wavelength range from $\rm \sim4000\AA $ to $\rm 9000\AA$. 
We have also observed the flux calibrator `Feige 110' on each night after Gaia 20eae to flux calibrate our HFOSC spectra. 
On 2020  September  12, we also obtained a medium resolution optical spectrum (R$\sim$1140, 1760 and 1920 for the Orange 
Arm, Red Arm and Far Red Arm respectively) using the LRS-2 Red
 Integral Field Unit spectrograph on 10-m Hobby-Eberly Telescope (HET) \citep{HETRamsey,HETqueue}, USA. LRS2-R spectrum was reduced using standard LRS2 pipeline, 
Panacea\footnote{\url{ https://github.com/grzeimann/Panacea/blob/master/README\_v0.1.md}}. Finally, we scaled 
our flux calibrated spectra  to match the flux and slope obtained from the photometric flux values of the same date. In case,
 photometric magnitudes are not available on the same
 date we have scaled our spectra with the
 photometric flux values of the nearest date. This
 is done to correct for any residual systematics in the flux calibration for the HFOSC due to its sensitivity to seeing variations and centering errors on the slit. This is also important for the LRS2 IFU observation, since the night was hazy due to smoke from wildfires, resulting in a highly variable non-grey atmospheric extinction.

We obtained NIR spectra of Gaia 20eae using the TANSPEC with its 0$^{\prime\prime}.5$ slit providing a R$\sim$2700 on the nights 
of 2020 October 24 and 2020 November 6. Standard NIR dithering technique i.e, obtaining the spectra 
at two different slit positions, was followed. The final
spectrum of the object is obtained by subtracting the spectra obtained at the dithered positions to cancel 
the sky contribution. A telluric standard star was also observed immediately after the Gaia 20eae observations to remove the telluric features.

We have used the standard tasks of IRAF\footnote{IRAF is distributed by National Optical Astronomy
Observatories, USA which is operated by the Association of Universities for Research in Astronomy, Inc., under cooperative agreement with National Science Foundation for performing image processing.}
to reduce medium resolution spectroscopic data.
The task {\sc apall} was used to extract the one dimensional spectrum. The extracted spectrum was then calibrated using {\sc identify} task  with 
the help of the calibration lamps taken immediately after the source spectrum.
Finally, {\sc continuum} task of IRAF is used to continuum normalize the spectra in order to measure the equivalent widths (EWs) of different lines. 
 Standard IRAF tasks
{\sc standard}, {\sc sensfunc} and {\sc calibrate} were used to flux calibrate our spectra.

\subsubsection{High Resolution Near Infrared Spectroscopy}

We observed a high resolution NIR spectrum of  Gaia 20eae on 2020 September 12 using the Habitable Zone Planet Finder (HPF) \citep{2012SPIE.8446E..1SM,mahadevan_habitable-zone_2014} on the  10m  HET.  HPF covers the wavelength range of 8100 -- 12800 $\AA$, at a spectral resolution of R$\sim$55,000. The H2RG up-the-ramp raw data cube was reduced to 1D spectra by the procedures described in \citet{ninan_habitable-zone_2018,kaplan2018,stefansson20}. The wavelength calibration was done using a laser frequency comb calibrator as described in \citet{2019OptL...44.2673M}. Barycentric correction was applied to all spectra with the values calculated using \texttt{barycorrpy}  \citep{kanodia_python_2018}.

In summary, we have monitored  Gaia 20eae spectroscopically at 10 different epochs with the HFOSC (6), TANSPEC (2), LRS2-R (1) and HPF (1). Table \ref{tab:obs_log} 
provides the complete log of spectroscopic observations presented in this work.

\section{Results and Analysis} \label{pt2}

\begin{figure*}
\centering
\includegraphics[width=0.95\textwidth]{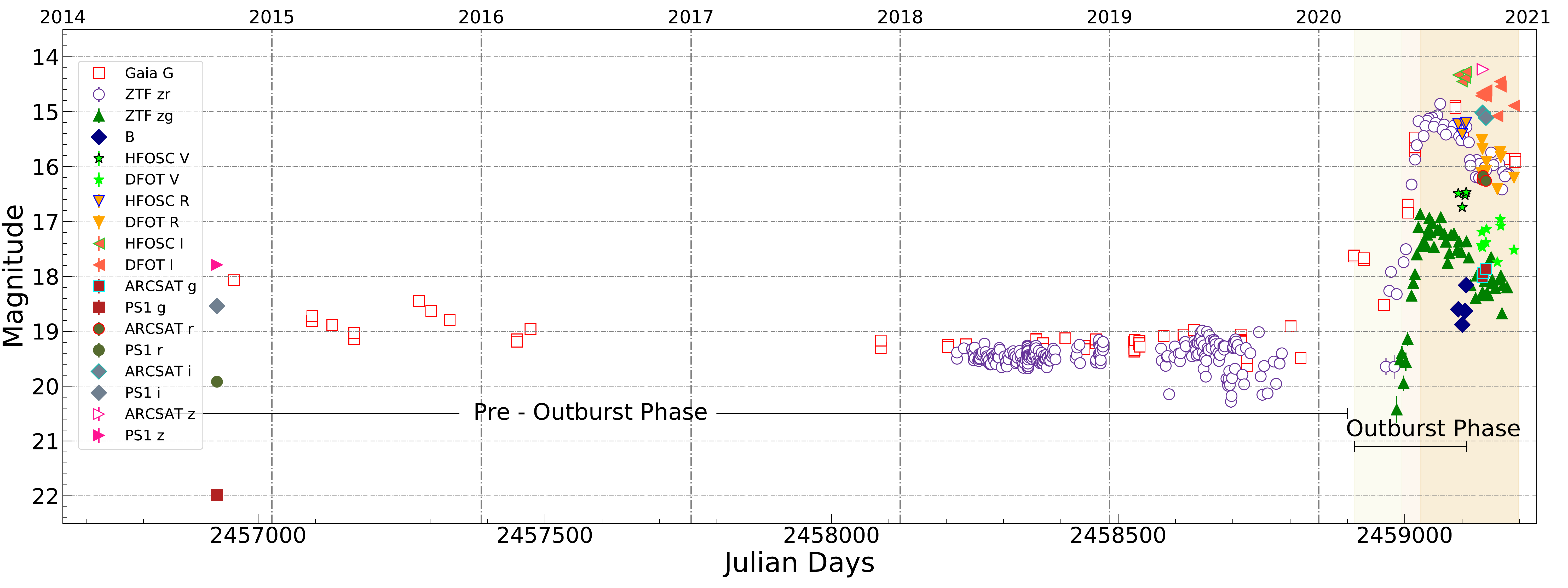}
\includegraphics[width=0.95\textwidth]{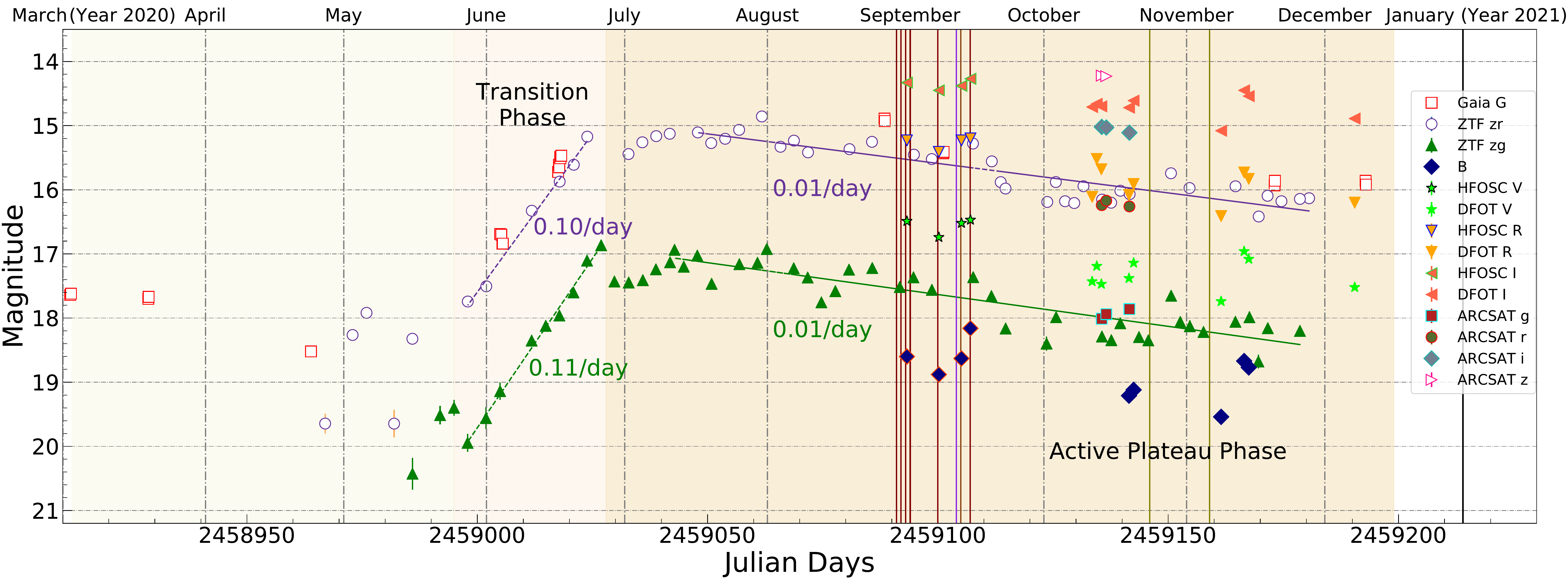}
\caption{\label{flc}Upper panel: Historical Light Curve (LC) of  Gaia 20eae in $Gaia$ $G$, ZTF $zg$ and $zr$, Johnson-Cousins $B$, $V$, $R$ 
and $I_{C}$ and $SDSS$ $g$, $r$, $i$ and $z$ bands showing both pre-outburst (white region) and outburst (shaded region) phases. 
Lower panel: Zoomed-in LC of  Gaia 20eae in the outburst phase.
The dark and intermediate brown regions represent the current plateau phase and the transition region, respectively.
The vertical brown and magenta solid lines are representing the epochs 
when HCT HFOSC and HET HPF spectra were taken, respectively. Olive line denotes the epoch of TANSPEC spectroscopic observations.
}
\end{figure*}

\subsection{Gaia 20eae during Quiescent phase}

\subsubsection{Physical properties}

Gaia 20eae is named as `SSTGLMC G050.2584-00.5077' and classified as a candidate young stellar source (YSO) 
due to its red color in MIR bands using the $Spitzer$ photometry by \citet{2008AJ....136.2413R}.
Later on, this source was classified as Class\,{\sc ii} YSO based on its IR spectral index \citep[MC1-M15 in the Table 4 of][]{2017ApJ...839..113R}.
\citet{2017ApJ...839..113R} also derived its mass as 1.5 M$_\odot$ assuming an age of 2 Myr for a typical Class\,{\sc ii} source and a distance of 1.3 kpc.
As \citet{2021AJ....161..147B} have estimated a distance of 3.2 kpc for Gaia 20eae from Gaia DR3, this would result in a different mass estimation.
However, in the absence of direct measurement of $A_V$ around this source, it would be very difficult to derive its accurate physical parameters (e.g., age/mass).

\subsubsection{Light Curve}

The upper panel of Figure \ref{flc} shows the Light Curve (LC) of Gaia 20eae in $Gaia$ $G$, ZTF $zg$ and $zr$, $Johnson-Cousins$
$B$, $V$, $R$ and $I$ and $SDSS$ $g$, $r$, $i$ and $z$ bands. 
It is worthwhile to mention here that, although the ZTF  \citep{Bellm_2018} and  $SDSS$  filters cover similar wavelengths,
they have differences in their cutoff wavelength and transmission curve.
The Gaia $G$ band data cover the longest time span of the LC starting from 2014 October 27 (JD=2456957) upto 
2020 December 9 (JD=2459192).
The ZTF $zr$ band has data from 2018 April 9 (JD$=$2458218) to 2020 November 27 (JD$=$2459180), whereas 
$zg$ band data are available from 2020 May 16 (JD$=$2458985) to 2020 November 25 (JD$=$2459178).
The ZTF photometric data have better temporal sampling (2-3 days) as compared to the  Gaia $G$ band data (10-15 days).
The LC clearly demonstrates a long quiescent period with minor fluctuations until 2019 October 28 (JD$=$2458785),
after that, it began to transit to the present outburst stage. 
The pre-outburst  magnitudes of  Gaia 20eae were  $G\sim19.49$ mag (2019 November 30; JD$=$2458818) and $zr\sim19.40$ mag (on 2019 October 28; JD$=$2458785). 
The mean $G$ and $zr$ magnitudes of Gaia 20eae during the quiescent phase were  $19.14\pm0.26$ mag (from 2014 October 27 to 2019 November 30) 
and $19.46\pm0.17$ mag (from 2018 April 9 to 2019 October 28), respectively. 

\begin{figure}
\centering
\includegraphics[width=0.45\textwidth]{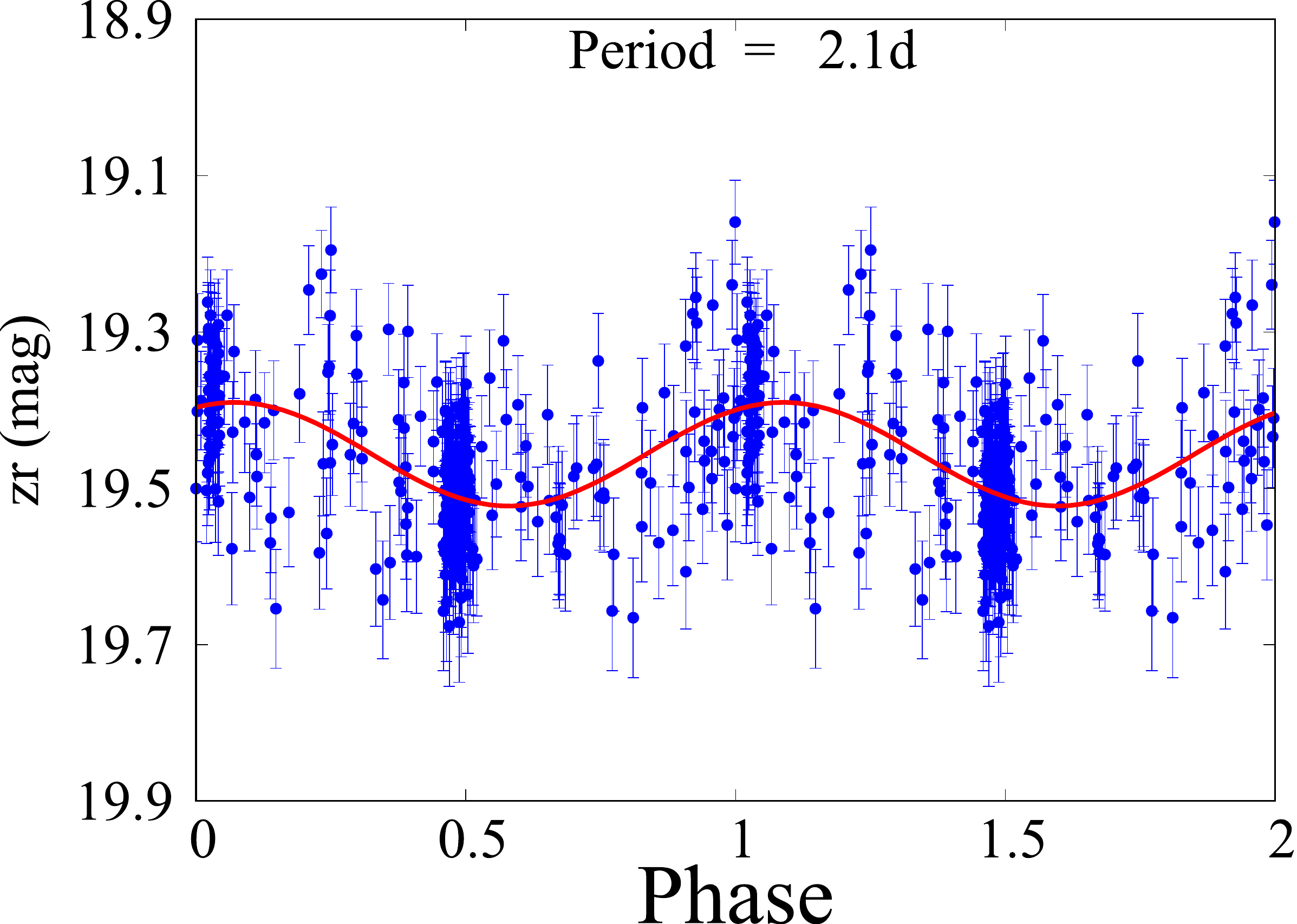} 
\vspace{0.5cm} 
\caption{\label{period_quiescent} Quiescent phase folded LC of Gaia 20eae as obtained from the ZTF $zr$ band data. The period is determined by 
using the Period software and also cross-matched with the NASA Exoplanet Archive Periodogram service.
}
\end{figure}

As the quiescent phase LC of Gaia 20eae shows small scale fluctuations, we searched for periodic variability in it using the ZTF $zr$ band data.
Periodic variability has been reported in the PMS stars which  is due to the rotation of the star having hot and cool spots on their photosphere.  
We have used the Period\footnote{\url{http://www.starlink.rl.ac.uk/docs/sun167.htx/sun167.html}} software, 
which works upon the principle of Lomb-Scargle (LS) periodogram \citep{1976Ap&SS..39..447L, 1982ApJ...263..835S}, to determine the period of
Gaia 20eae and to phase fold the LC. The advantage of the LS method is that it is effective even in 
case of the data set being non-uniformly sampled. We have also used the NASA Exoplanet Archive Periodogram\footnote{\url{https://exoplanetarchive.ipac.caltech.edu/docs/tools.html}} service for cross verification.
The periods obtained in both the cases matched well. The period of Gaia 20eae thus comes out to be 2.1$\pm$0.004 days.
The period detected in the quiescent LC of the Gaia 20eae might correspond to the rotational period of the star.
This type of period is commonly observed in Class\,{\sc ii/iii} type of YSOs as shown by \citet{2020MNRAS.493..267S}. Figure \ref{period_quiescent} shows the phase folded 
LC of the  Gaia 20eae during its quiescent phase. The amplitude of variation is of the order of 0.2 mag which is also typical of 
Class\,{\sc ii/iii} type of YSOs \citep{2020MNRAS.493..267S}.

\begin{figure*}
\centering
\includegraphics[width=0.46\textwidth]{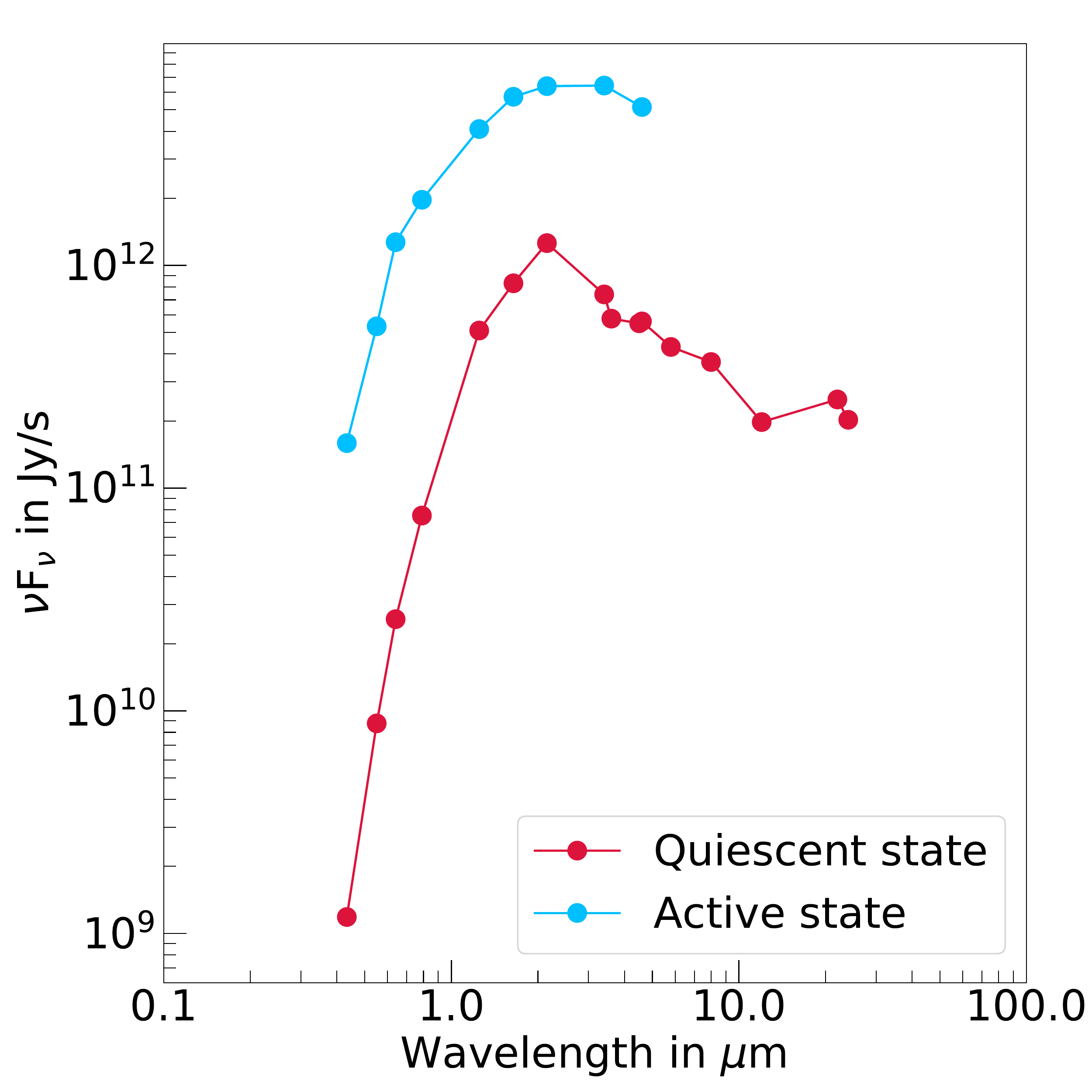}
\includegraphics[width=0.45\textwidth]{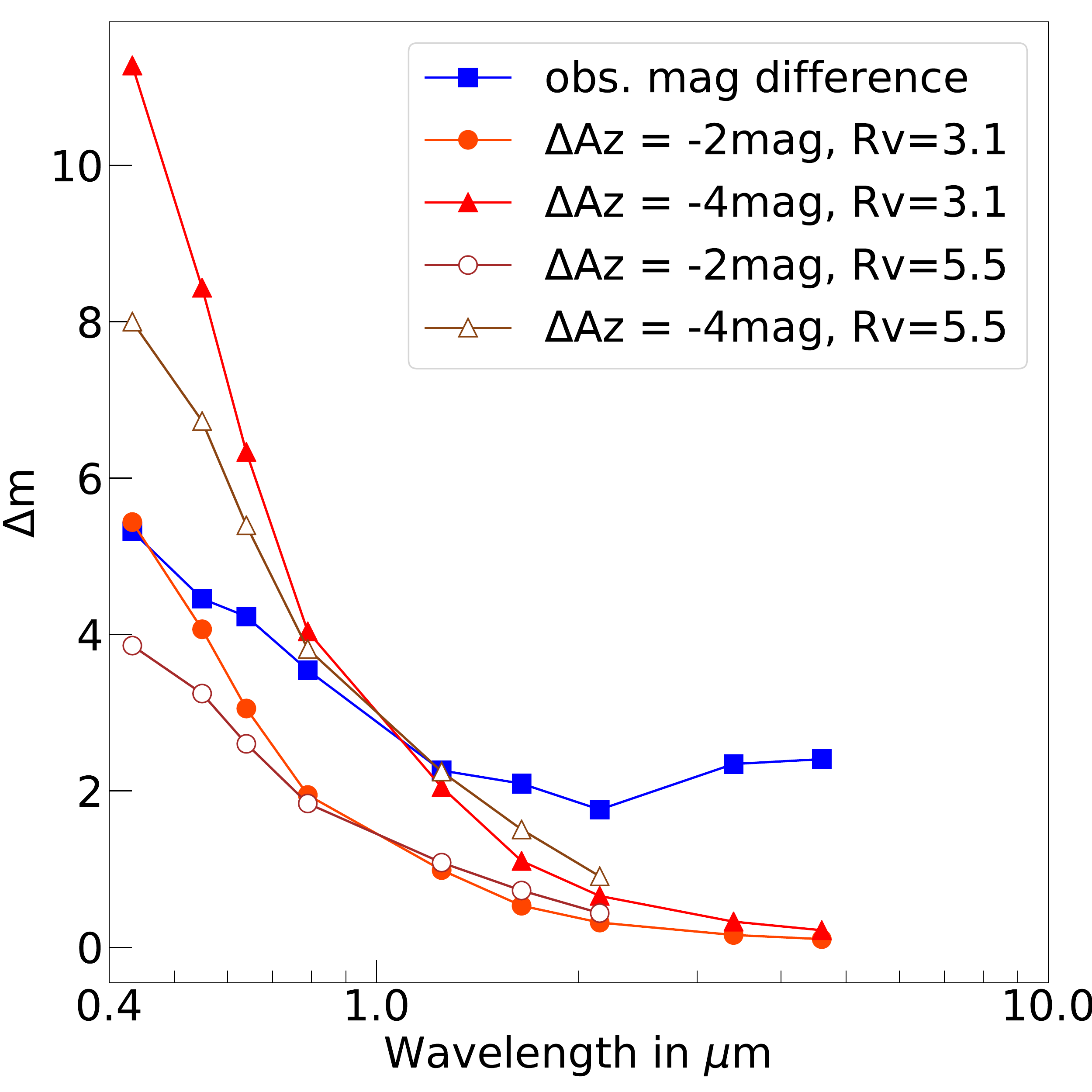}
\caption{\label{SEDnew}Left panel: The photometric SEDs of Gaia 20eae in quiescent (red dots) and 
active (cyan dots) phases. Right panel: Changes in SEDs, the observed magnitude
differences (squares) are compared against pure dust-clearing events for
$A_z = -2$ mag (circles) and $A_z= -4$ mag (triangles) and for $R_V$=3.1
(filled) or $R_V$=5.5 (open) dust laws. }
\end{figure*}

\begin{table*}
\centering
\caption{Reddening invariant colors in the quiescent and the outburst phases of the Gaia 20eae. N$\sigma$ is the ratio of color change to the quadrature
	sum of measurement errors.}
\label{tab:cc_table1}
\begin{tabular}{@{}cccccccccc@{}}
\hline 
Days & $Q_{BVR}\pm\sigma$ & N$\sigma$ & $Q_{VRI}\pm\sigma$& N$\sigma$ & Days & $Q_{BVR}\pm\sigma$& N$\sigma$ & $Q_{VRI}\pm\sigma$ & N$\sigma$\\
(JD) & (mag)     &           &  (mag)   &           & (JD) & (mag)    &           &   (mag)   & \\       
\hline
2456928 & 0.25$\pm$0.15 & $-$ & 0.00$\pm$0.07 & $-$    & 2459135 &  $-$              & $-$ & 0.88$\pm$0.02 & $-12.0$\\
2459093 & 0.18$\pm$0.04 & 0.5 & 0.42$\pm$0.01 & $-5.9$ & 2459141 & $-0.16$ $\pm$0.04 & 2.7 & 0.03$\pm$0.02 & $-0.5$\\
2459100 & 0.10$\pm$0.01 & 1.0 & 0.44$\pm$0.01 & $-6.3$ & 2459142 & 0.09$\pm$0.05     & 1.0 & 0.02$\pm$0.03 & $-0.3$\\
2459105 & 0.14$\pm$0.01 & 0.8 & 0.50$\pm$0.01 & $-7.0$ & 2459161 & $-0.24$ $\pm$0.04 & 3.3 & 0.09$\pm$0.02 & $-1.3$\\
2459107 & $-0.25$ $\pm$0.01   & 3.4 & 0.40$\pm$0.01 & $-5.7$ & 2459166 & $-0.17$ $\pm$0.01 & 2.8 & 0.04$\pm$0.01 & $-0.5$\\
2459133 &  $-$                & $-$ & 0.02$\pm$0.02 & $-0.2$ & 2459167 & $-0.22$ $\pm$0.04 & 3.1 & 0.05$\pm$0.01 & $-0.7$\\
2459134 &  $-$                & $-$ & 0.87$\pm$0.02 & $-11.9$& 2459190 & $-$               & $-$ & 0.10$\pm$0.01 & $-1.4$\\
\hline 
\end{tabular}
\end{table*}

\subsection{Gaia 20eae during outburst phase}

\subsubsection{Light Curve}

In the lower panel of Figure \ref{flc}, we show the LC of Gaia 20eae in $B$, $V$, $R$, $I$, $G$, $zg$, $zr$, $g$, $r$, $i$, $z$ bands in the outburst phase. 
The LC starts from 2020 March 3 (JD $=$ 2458800) upto the latest data point on 2020 December 9 (JD $=$ 2459192).
The LC of Gaia 20eae is peculiar in the sense that the rise to peak brightness consists of two parts: an initial slow rise from the
quiescent phase starting from JD $=$ 2458800 to JD $=$ 2458995 and then a rapid rise to the peak brightness from JD $=$ 2459014 to JD $=$ 2459047, 
reaching maximum brightness on JD $=$ 2459047, and then a slowly decaying  phase (JD $=$ 2459047 to JD $=$ 2459145).
It also shows  small scale fluctuations with amplitude of $\sim$0.2 mag on a time scale of few days.
We were not able to derive the periodicity of  these fluctuations using the LS periodogram.
We call the rapid rise part and slow decaying part of the LC as transition phase and active plateau phase, respectively,
and the same have been labeled and shadowed  with different colors in the lower panel of Figure \ref{flc}.
We have calculated the rise-rate and decay-rate of Gaia 20eae in different wavelengths from the available Gaia and ZTF data 
by fitting a least square straight line in the different phases of the LC.
The fits for the data points  in the transition and active plateau phases are shown in the lower panel of  Figure \ref{flc}. 
The overall best-fit rise-rate from the quiescent phase to the maximum brightness is calculated to be 0.6 mag/month in the $G$ band,
over a duration of $\sim$247 days. This rise rate is prone to higher uncertainties as there are lots of data gaps in the LC initially.
The rise-rate in transition phase was found to be similar i.e. $\sim$0.1 mag day$^{-1}$ or $\sim$3 mag month$^{-1}$ in $G$, $zg$ and $zr$ bands.
The decay rate in the  active plateau phase is calculated as 0.01 mag day$^{-1}$ (or 0.3 mag month$^{-1}$)
for a duration of $\sim$98 days, which is an order less than the rise-rate. It is to be noted here that Gaia 20eae has not returned 
to its quiescent state. Hence, the decay rate that we calculated is by considering the data range that we have presented in this study. 
 
The maximum brightness in the current outburst phase in the $zg$ and $zr$ bands is recorded on 2020 July 16
(JD$=$2459047) as 17.02 mag and 15.09 mag, respectively.
In the $G$ band, the source reached the maximum brightness of 14.89 magnitude on 2020 August 26 (JD $=$ 2459088).
Thus, the current outburst magnitude amplitudes are: $\Delta G$ $=$ 4.25 mag and $\Delta zr$ $=$ 4.37 mag.  

   The LCs in $B,V,R$ and $I$ also follow the trend of the ZTF and Gaia LCs. The quiescent phase $J,H,$ and $K_s$ magnitudes of Gaia 20eae are
14.67 mag, 13.36 mag and 12.16 mag, respectively as obtained from the UKIDSS DR10plus. During the present outburst stage,
the $J,H,$ and $K_s$ magnitudes of Gaia 20eae as obtained from TANSPEC are 12.41 mag, 11.27 mag and 10.40 mag, respectively.
This implies that the present outburst is similar to the FUor family of objects and is almost wavelength independent.

\subsubsection{The evolution of the photometric Spectral Energy Distribution}

Figure \ref{SEDnew} left panel represents the Spectral Energy Distributions (SEDs) of Gaia 20eae during its quiescent and active states
represented in red and cyan curves, respectively. We constructed the quiescent phase SED using the multiwavelength data 
(optical to MIR wavelengths, i.e. 0.44 ($B$), 0.55 ($V$), 0.65 ($R$), 0.80($I$), 1.2 ($J$), 1.6 ($H$), 2.2 (K$_s$), 3.4($W1$), 
3.6($I1$), 4.6($W2$), 4.5($I2$), 5.8($I3$), 8.0($I4$), 12($W3$), 22($W4$) and 24($I4$) $\mu m$) taken from
the data archives (PS1, 2MASS, $Spitzer$ and $WISE$). 
The PS1 magnitudes were transformed to Johnson Cousins system by using the equations given by \citet{2012ApJ...750...99T}. 
For the  outburst state SED, we have used current optical and NIR band observations as well as NEOWISE data. 
Apart from having a shift in the brightness, there is clearly a change in the shape of the SED at the longer 
wavelengths as compared to the shorter wavelengths. This change in the SED can be quantified by
comparing the differences in observed magnitudes of Gaia 20eae with that of pure dust clearing events. 
Right panel of Figure \ref{SEDnew} shows the observed magnitude differences of Gaia 20eae between its 
quiescent and active state by a blue curve. We compare the magnitude variations 
against $\Delta$A$_z$ = $-2$ mag and $\Delta$A$_z$ = $-4$ mag
for R$_V$ = 3.1 and R$_V$ = 5.5 dust laws \citep[see also,][]{2004ApJ...616.1058M}.  
From the deviation of the observed magnitude difference with the dust clearing events, we can conclude that the brightening of Gaia 20eae cannot be 
explained by the diminishing of the line-of-sight extinction, rather the enhancement of accretion rate is the likely cause.

Using the multi-epoch data from the PS1 archive (quiescent phase), 2m HCT and 1.3m DFOT (outburst phase), 
we have also examined the evolution of the reddening invariant colors, `$Q_{xyz}$' of Gaia 20eae as it transitioned from quiescent state to the eruptive state. 
The reddening invariant colors have a generic form of :  
   $Q_{xyz} = (x-y)-[(y-z)E(x-y)/E(y-z)]$,
where x, y, and z are the observed magnitudes in each passband  \citep{2004ApJ...616.1058M}.  
A color change having $\Delta$ $Q_{xyz}$ $\neq$ 0 indicates that the changes in SED is not due to pure dust clearing. 
The estimated reddening-invariant colors  (for  R${_V}$ = 3.1) for Gaia 20eae as it transitioned from the quiescent state 
to the active state are listed in Table \ref{tab:cc_table1}. 
The large change in most of the Q$_{xyz}$ values also points towards that the increase in the brightness of Gaia 20eae is not consistent with a dust-clearing event, 
rather an intrinsic change occurred in the SED. It is also to be noted that in the `Active Plateau Region' of Gaia 20eae starting from JD$=$2549141,
the value of Q$_{VRI}$ is close to 0.  This might imply that there was no change in the SED intrinsically in this duration.

\begin{figure*}
\centering
\includegraphics[width=0.98\textwidth]{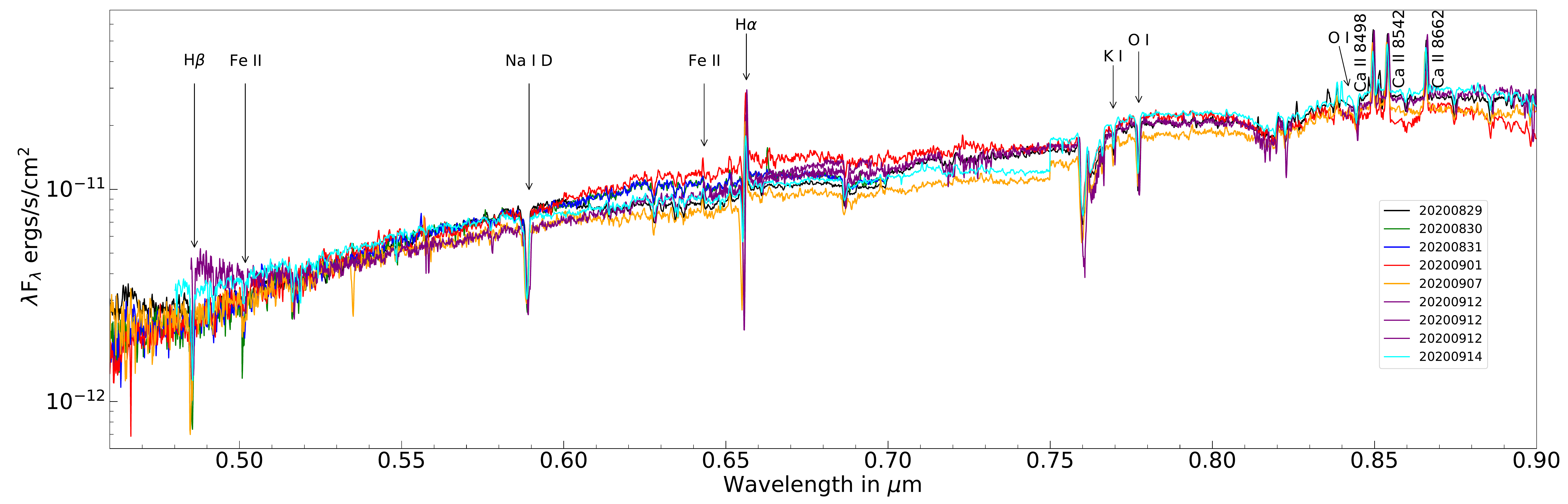}
\includegraphics[width=0.95\textwidth]{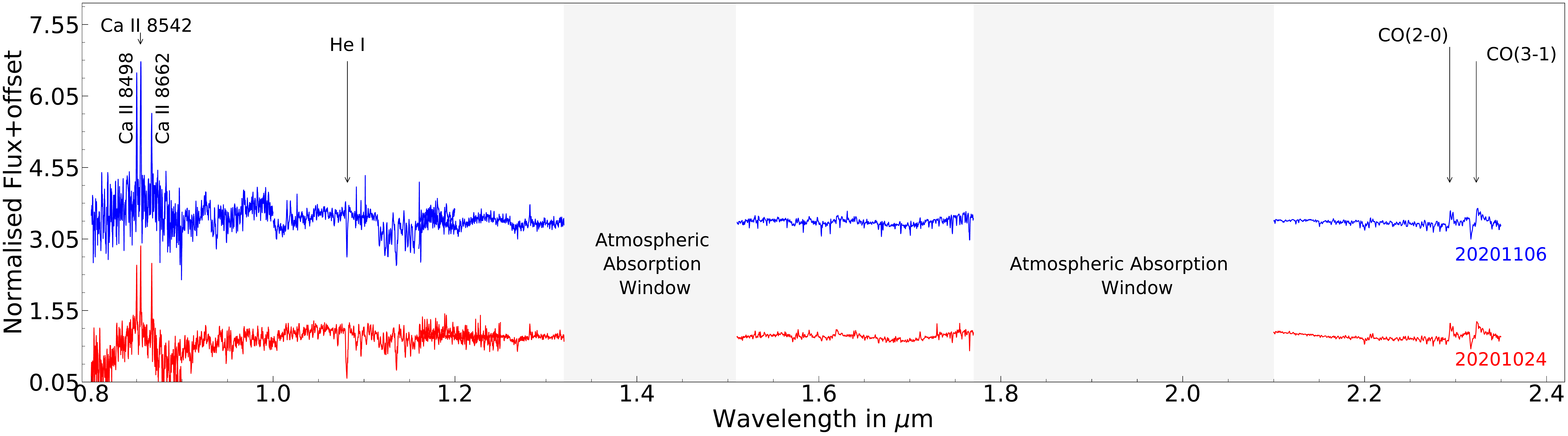}
\caption{\label{frr} Evolution of the flux-calibrated spectra of Gaia 20eae during our monitoring period obtained using HFOSC on 2m HCT and
LRS2 on 10m HET starting from 2020 August 29 to 2020 September 14 (upper panel). The lines used for the present 
study have been marked. The flux$-$calibrated spectrum show variation in the continuum level which is also present in the LC. Lower panel shows
the normalized flux of Gaia 20eae obtained using TANSPEC on 2020 October 24 and 2020 November 06.}
\end{figure*}

\subsubsection{Spectral features}

\begin{figure*}
\centering
\includegraphics[width=0.95\textwidth]{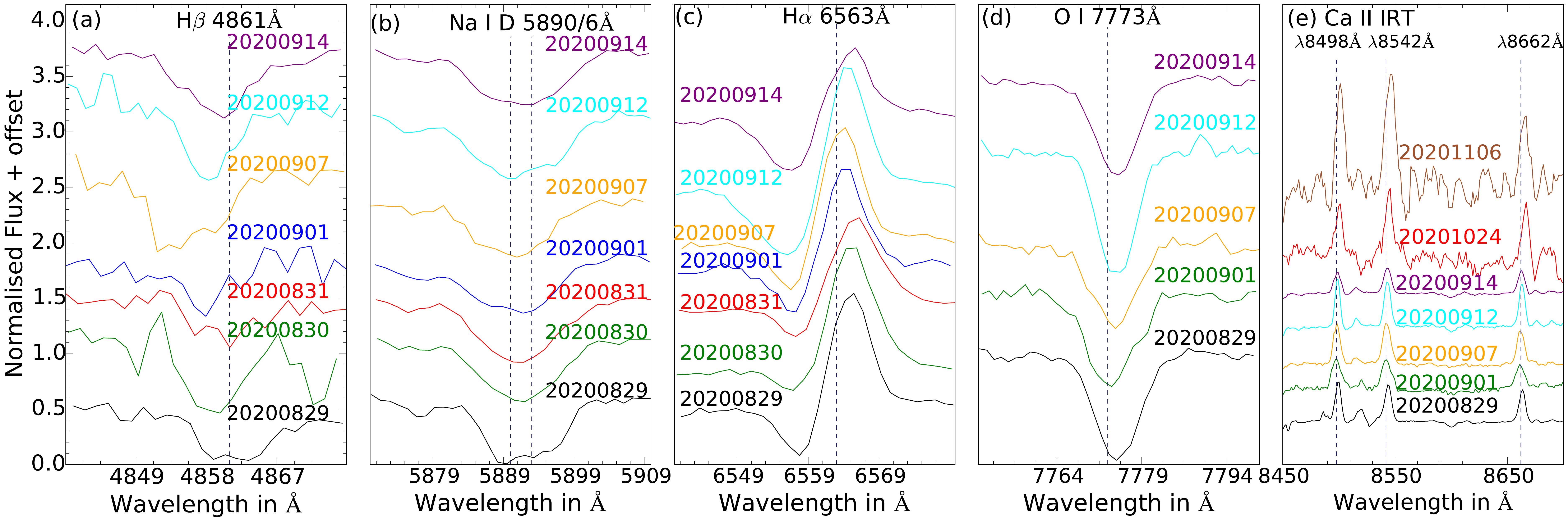}
\includegraphics[width=0.95\textwidth]{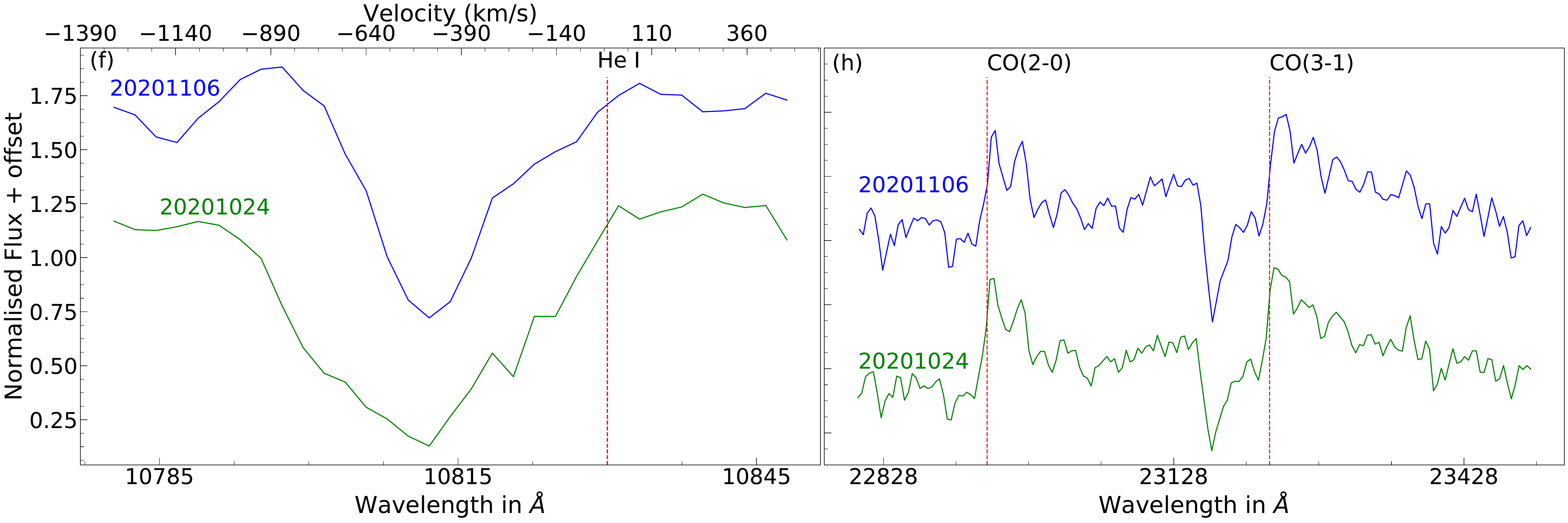}
\caption{\label{fca} Evolution of the H$\beta$, Na I D $\lambda$5890/6$\rm\AA$, H$\alpha$, O I $\lambda$7773$\rm\AA$, Ca II IRT, He I, CO(2-0) and CO(3-1)
lines during our monitoring 
period.}
\end{figure*}

Figure \ref{frr} shows our medium resolution spectra covering a very broad wavelength range ($\sim$0.4 - 2.4 $\mu$m).
The optical spectra are flux-calibrated whereas the NIR spectra are only normalized  spectra.
Both the optical and NIR spectra of Gaia 20eae consist of a mixture of the lines typically observed in the FUor and EXor family of sources. 
Evolution of these spectral lines at different epochs of the outburst phase are shown in 
Figure \ref{fca}. 
Similar to the FUor sources,
Gaia 20eae exhibits  blue-shifted absorption features in Na I resonance line and H$\beta$ line, indicative of the powerful winds
from the source. It also shows strong P Cygni profile in H$\alpha$ and Ca II IR triplets (IRT) in emission. 
Fe II line at $\lambda$5018$\rm\AA$ which,
is seen in emission in EX Lupi, is found to be in absorption but the Fe II line at $\lambda$6433$\rm\AA$ is found to be in emission. 
 K I $\lambda$7694$\rm\AA$ and O I $\lambda$8446$\rm\AA$ lines are found to be in absorption. 
The strength of H$\beta$ and Na I D lines can be seen decreasing during the outburst phase of Gaia 20eae.
Spectroscopically, in the optical regime the spectrum of Gaia 20eae resembles 
 a FUor, whereas in the NIR regime it is more or less similar to an EXor.

Our medium resolution NIR spectra show several distinct spectral features, most of them are in absorption. 
The gaps in the spectra represent atmospheric absorption windows due to the broad H$_2$O and OH bands.
We could identify some prominent lines: Ca II IRT, He I at $\lambda$10830$\rm\AA$ 
and the CO bandheads. 

\textbf{CO bandhead in K band:} The CO (2-0) and CO(3-1) bandhead absorption are one of the defining characteristics of FUors \citep{1998apsf.book.....H}. The CO bandheads in Gaia 20eae are in emission, implying a temperature inversion at the surface of the protoplanetary disc. This is similar to  that observed in other EXor sources like V2492 Cyg \citep{2011AJ....141..196A}. Thus based on the CO bandhead lines, Gaia 20eae resembles more of an EXor source.\footnote{It should be noted that some intermediate FUor/EXor sources like V1647 Ori have shown CO bands in both emission as well as in absorption at different stages of its outbursts.}

\textbf{Radial velocity of Gaia 20eae:} Due to lack of symmetric photospheric lines in the high resolution spectrum of Gaia 20eae, it is hard to estimate the radial velocity of the star. The chromospheric Fe I emission lines were found to be the most symmetric lines in the spectrum, and the line center of the Fe $\lambda$8387.77$\rm\AA$ at 20 km s$^{-1}$ is taken as the stellar radial velocity 
with respect to the solar system barycentre in this study. The corresponding velocity in the local standard of rest reference frame comes out to be $\sim$35 km s$^{-1}$.
It is to be mentioned that the peak velocity of the $^{13}$CO for the molecular cloud `MC2' is 42 km s$^{-1}$ with respect to the local standard of rest \citep{2017ApJ...839..113R}.

\subsection{Physical parameters of Gaia 20eae during outburst phase}

\subsubsection{Disc turbulence and outflow wind velocities}

The panel (d) in Figure \ref{fca} represents the evolution of O I line at $\lambda$7773$\rm\AA$ of Gaia 20eae during our spectroscopic monitoring period. 
The formation of O I line at $\lambda$7773$\rm\AA$ in T Tauri stars, which is an indicator of turbulence, is attributed to the presence of warm gas 
in the envelope surrounding the disc or in the hot photosphere above the disc \citep{1992ApJS...82..247H}. Table \ref{tab:velocity} shows the
variation in the value of equivalent width (EW) of O I line at $\lambda$7773$\rm\AA$ during our monitoring period. 
The mean value of the equivalent width of O I line at $\lambda$7773$\rm\AA$ is estimated 
as $4.7 \pm 1.2$ $\rm\AA$.
The scatter in the EW values is twice the error in its estimation, indicating the presence of turbulent medium surrounded around Gaia 20eae
during its outburst phase. 
For future inference, we have also tabulated the EW values of other lines during different epochs of the outburst phase of Gaia 20eae in Table  \ref{tab:velocity}.

Outflow wind velocity of the Gaia 20eae is estimated from the blue-shifted absorption minima of H$\alpha$, Na I D and H$\beta$ lines 
\citep{1998apsf.book.....H}.
The estimated values of wind velocity by Doppler shift at different epochs in the outburst phase of Gaia 20eae are listed in the Table  \ref{tab:velocity}.
The values show the variation from $-630$ to $-203$ km s$^{-1}$.
The mean velocity of the outflow wind velocity for H$\alpha$, Na I D and H$\beta$ comes out to be $-505\pm62$ km s$^{-1}$, $-356\pm49$ km s$^{-1}$ and $-339\pm136$ km s$^{-1}$, respectively.
The typical error in the outflow wind velocity estimation is $\sim$25 km s$^{-1}$, 
therefore, a large scatter in its values can be attributed to intrinsic variation of outflow winds during the outburst  phase. 

\begin{figure}
\centering
\includegraphics[width=0.5\textwidth]{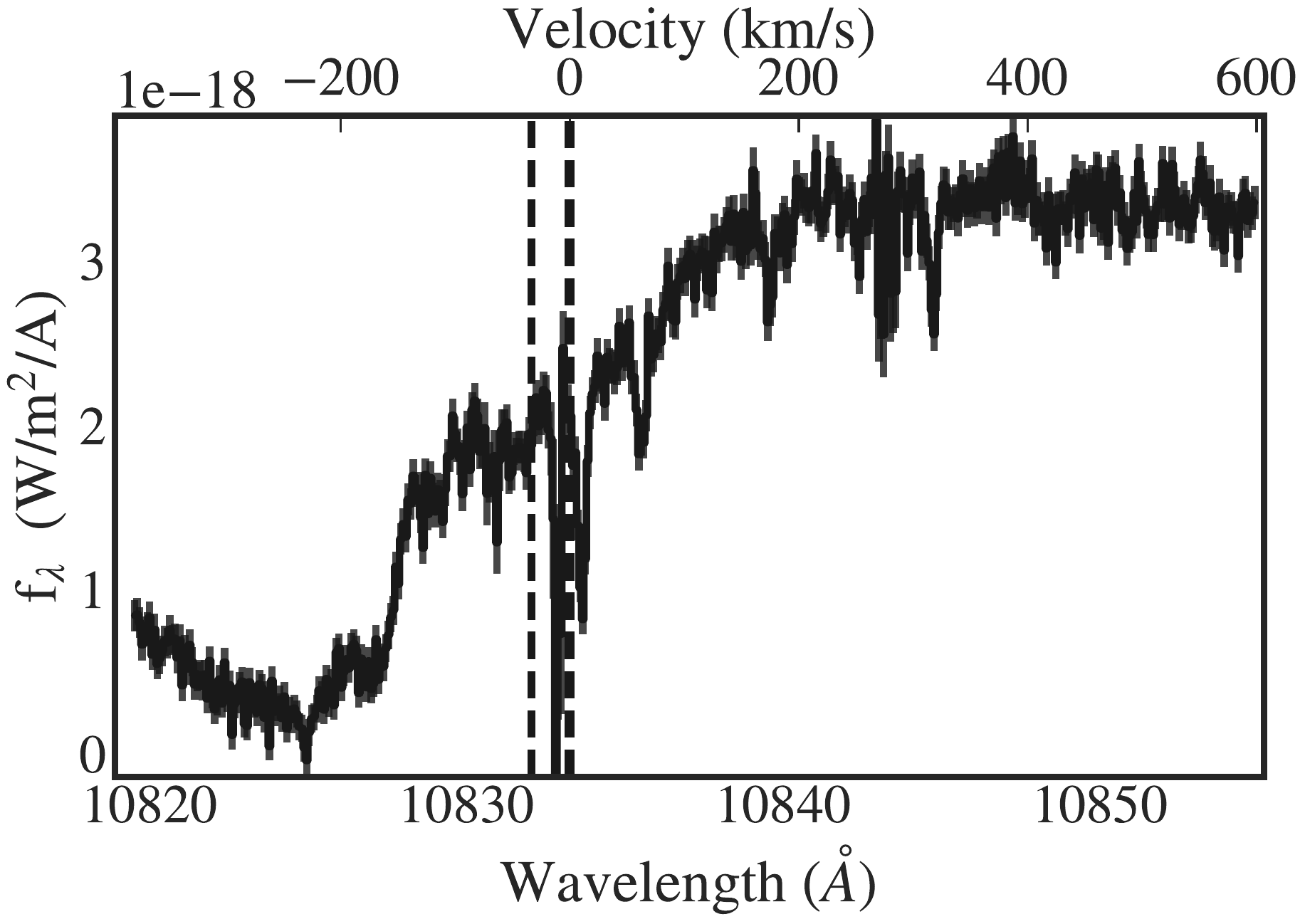}
\caption{\label{HPFHe10830} High resolution line profile of the He 10830 triplet on 2020 September 12 showing a large blue-shifted component, likely originating in the strong outflow from Gaia 20eae. The absorption signal extends from +200 km s$^{-1}$ on the red side, to beyond $-400$ km s$^{-1}$ on the blue side (the profile is truncated at the detector edge in HPF). The vertical dashed lines show the location of the He 10830 triplets in the stellar rest frame. The narrow lines in the spectrum are telluric.}
\end{figure}

Resonant scattering from meta-stable Helium atoms are excellent traces of the outflow winds from YSOs \citep{Edwards_2003}. The EUV to X-ray 
radiation from magnetospheric accretion or chromosphere activity can cause significant formation of meta-stable triplet ground state of Helium
atoms. During the $\sim$2.5 hours when they typically survive in this meta-stable state, they could resonant scatter the $\lambda$10830$\rm\AA$ photons resulting 
in a strong absorption signal at the local velocity of the gas. Figure \ref{HPFHe10830} shows the very strong blue-shifted He $\lambda$10830$\AA$
absorption signature in the high resolution spectrum of  Gaia 20eae. On the red side, the absorption profile extends to about +200 km s$^{-1}$ and on the blue side, the absorption 
profile extends beyond $-400$ km s$^{-1}$. Unfortunately, the high resolution spectrum beyond $-400$ km s$^{-1}$ could not be measured since it falls outside 
the detector in the HPF spectrograph. Our medium resolution TANSPEC spectra in the panel (f) in Figure \ref{fca} shows the blue-shifted absorption extending to $-513$ km s$^{-1}$ on 2020
October 24 and reducing to $-493$ km s$^{-1}$ by 2020 November 6. Such strong blue-shifted He $\lambda$10830$\rm\AA$ 
triplet signatures are common in YSOs
with strong outflows.  Gaia 20eae is not an exception. The reduction in the blue-shifted wing velocity of He $\lambda$10830$\rm\AA$ over a span of two weeks 
could be either due to change in the  structure of the outflow winds or due to drop in the EUV - X-Ray irradiation.

\subsection{Magnetospheric accretion and line profiles}

High resolution line profile shapes provide us direct measurement of the kinematics of the hot gaseous environment of the accretion region. In this section we highlight some of the most interesting line profiles we detected in  Gaia 20eae.

{\subsubsection{Infall signature in Ca II IRT} }

\begin{figure}
\centering
\includegraphics[width=0.45\textwidth]{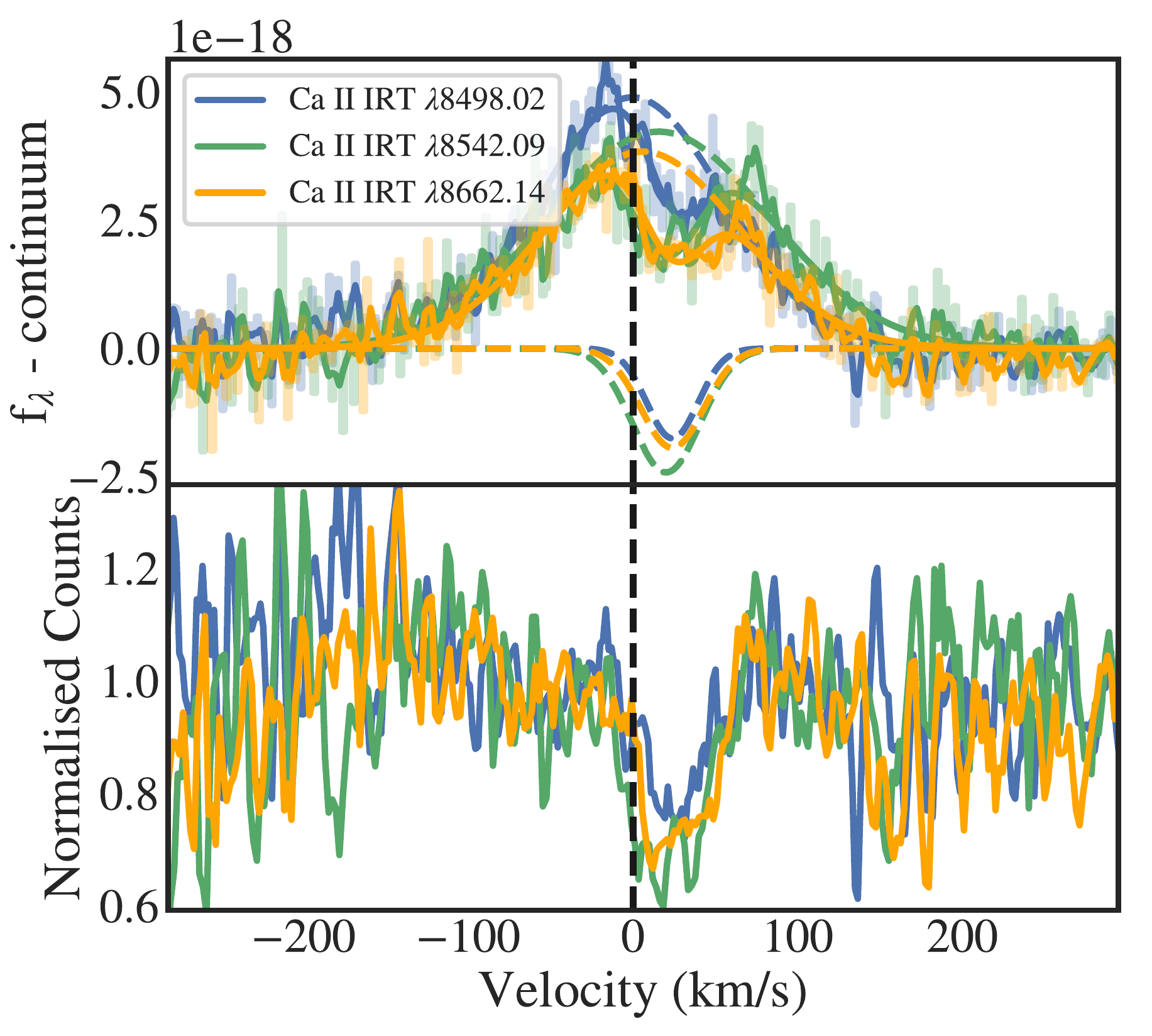}
\caption{\label{HPFCaII} Top-panel: High resolution line profiles of all the three Ca II IRT lines on 2020 September 12 showing a red-shifted absorption component. This is likely originating in the magnetospheric accretion funnel. The fainter step style plots behind the bold lines are the measured spectra. The bold curves are a 3 pixel smoothed spectra shown for clarity. The thin smooth curve is the double Gaussian composite fit of the emission at the stellar rest velocity and a red-shifted absorption at $\sim$+25 km s$^{-1}$. The individual components are shown in dashed curves. 
Lower panel: Normalized spectrum using the continuum plus the fitted emission Gaussian component.}
\end{figure}

Ca II IR triplet emission lines are believed to originate in the active chromosphere as well as in the magneto$-$spheric accretion funnel regions
\citep{hamann1992,1998AJ....116..455M}. Our observation of Gaia 20eae on 2020 September 12 detected a red-shifted absorption component
(with respect to the stellar rest frame) in all the three Ca II IR triplet lines (See Figure \ref{HPFCaII}). The smooth curves show the best 
fit of a double Gaussian model, where the first component fits the broad emission line and the second component fits the red-shifted 
absorption component at +25 km s$^{-1}$. The red-shifted absorption cannot originate in stellar wind or outflows. One possible region of origin could
be the hot in-falling gas in the magnetospheric accretion funnel \citep{edwards1994}. The lower panel in Figure \ref{HPFCaII} shows this 
absorption component normalized to the fitted emission line profile. The ratio of equivalent widths of these absorption components (EqW: 0.27, 0.49, 0.41 $\AA$) 
is inconsistent with the optically thin line formation scenario (1:9:5).\footnote{This is unlike the optically thin blue-shifted absorption from winds typically 
seen in similar FUors/EXors V1647 Ori and V899 Mon \citep{2013ApJ...778..116N,2015ApJ...815....4N}.} 

\begin{figure}
\centering
\includegraphics[width=0.45\textwidth]{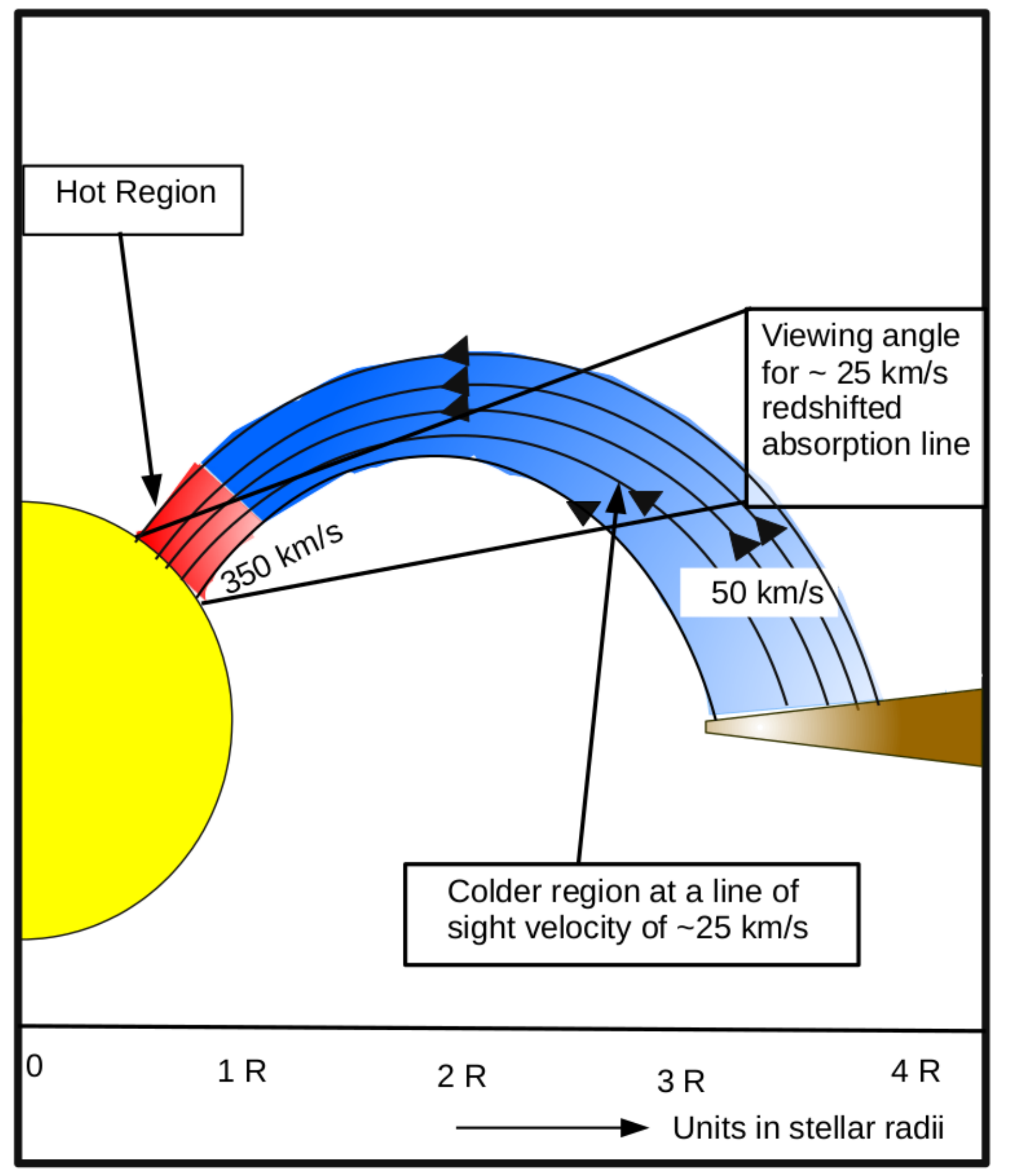}
\caption{\label{AccDiagram}  Diagram of the classical dipole magnetospheric accretion funnel, and the region of the viewing angles that could potentially result in a low velocity red-shifted absorption signature on top of the broad Ca II IR triplet emission line is shown in the diagram.}
\end{figure}

\textbf{Constrain on the viewing angle:} The detection of the absorption profile  implies the line of sight is along the increasing temperature gradient, and we are not seeing an infalling hot gas in the foreground of a cooler environment.  i.e., the viewing angle is through the accretion funnel to the footprint on the
stellar surface where it is hottest. For a star of mass M, radius $R_*$, and the disc infall radius $r_m$, the velocity of the infalling gas along the magnetic 
field line direction is given by the formula $v_p(r) = [\frac{2GM}{R_*} (\frac{R_*}{r} - \frac{R_*}{r_m})]^{1/2}$ \citep{hartmann1994}. For 
Gaia 20eae, this would imply a velocity of $\sim$350 km s$^{-1}$ at the base of the funnel, and in the order of $\sim$50 km s$^{-1}$ (or $\sim$25 km s$^{-1}$ 
along line of sight) at a radius very close to start of infall near the truncated accretion disc. This combined with the requirement of hotter
background against which this low velocity infall gas is viewed, constrains the viewing angle as shown in Figure \ref{AccDiagram}.

\subsubsection{Hydrogen Paschen lines}
\begin{figure}
\centering
\includegraphics[width=0.45\textwidth]{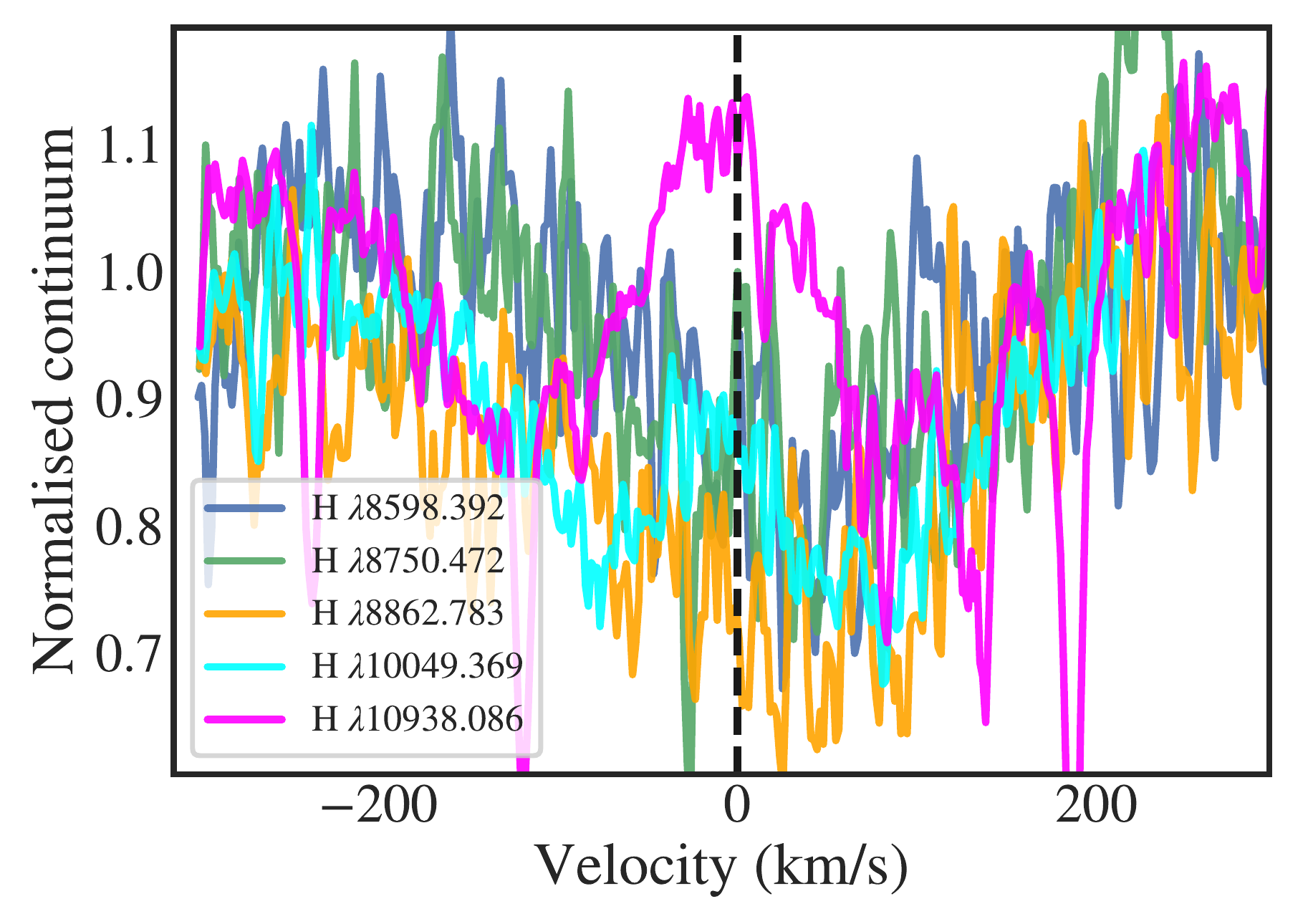}
\caption{\label{HPFHPa} Hydrogen Paschen lines from Pa (14-3) to Pa $\gamma$ (6-3). The lines which are completely lost in telluric bands are not plotted. The narrow lines in Pa$\gamma$ $\lambda 10938.086$ are telluric. The higher energy Paschen lines are detected as broad absorption lines starting from $-250$ km s$^{-1}$ to +250 km s$^{-1}$.  We detect emission at the line core for the lower energy level lines Pa$\gamma$ $\lambda 10938.086$ and Pa$\delta$ $\lambda 10049.369$ on top of the broad absorption component.   
The bold curves are a 3 pixel smoothed spectra shown for clarity.}
\end{figure}

Figure \ref{HPFHPa} shows the Hydrogen Paschen lines from Pa (14-3) to Pa $\gamma$ (6-3). Only the lines which are not completely lost in telluric bands are plotted here. The higher energy level Paschen lines are detected as broad absorption lines extending from $-250$ km s$^{-1}$ to +250 km s$^{-1}$. However, in the lowest energy levels lines, Pa$\gamma$ $\lambda 10938.086$ and Pa$\delta$ $\lambda 10049.369$, on top of the broad absorption component, we also detect an emission component at the core of the lines. The strength of this emission component decreases as we go to lines of higher energy levels.

\subsubsection{Fe I and Ti I lines}
\begin{figure}
\centering
\includegraphics[width=0.45\textwidth]{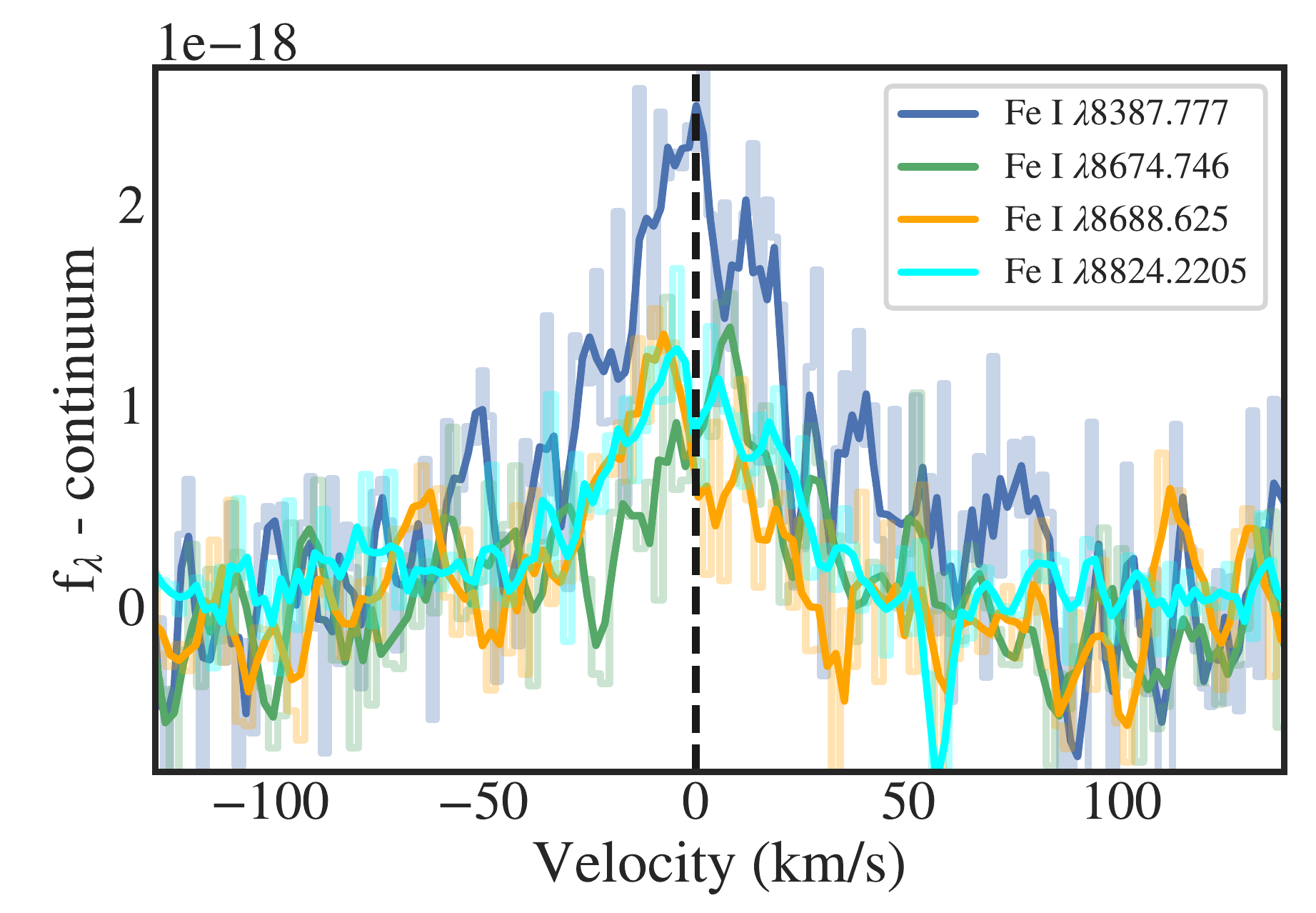}\\
\includegraphics[width=0.45\textwidth]{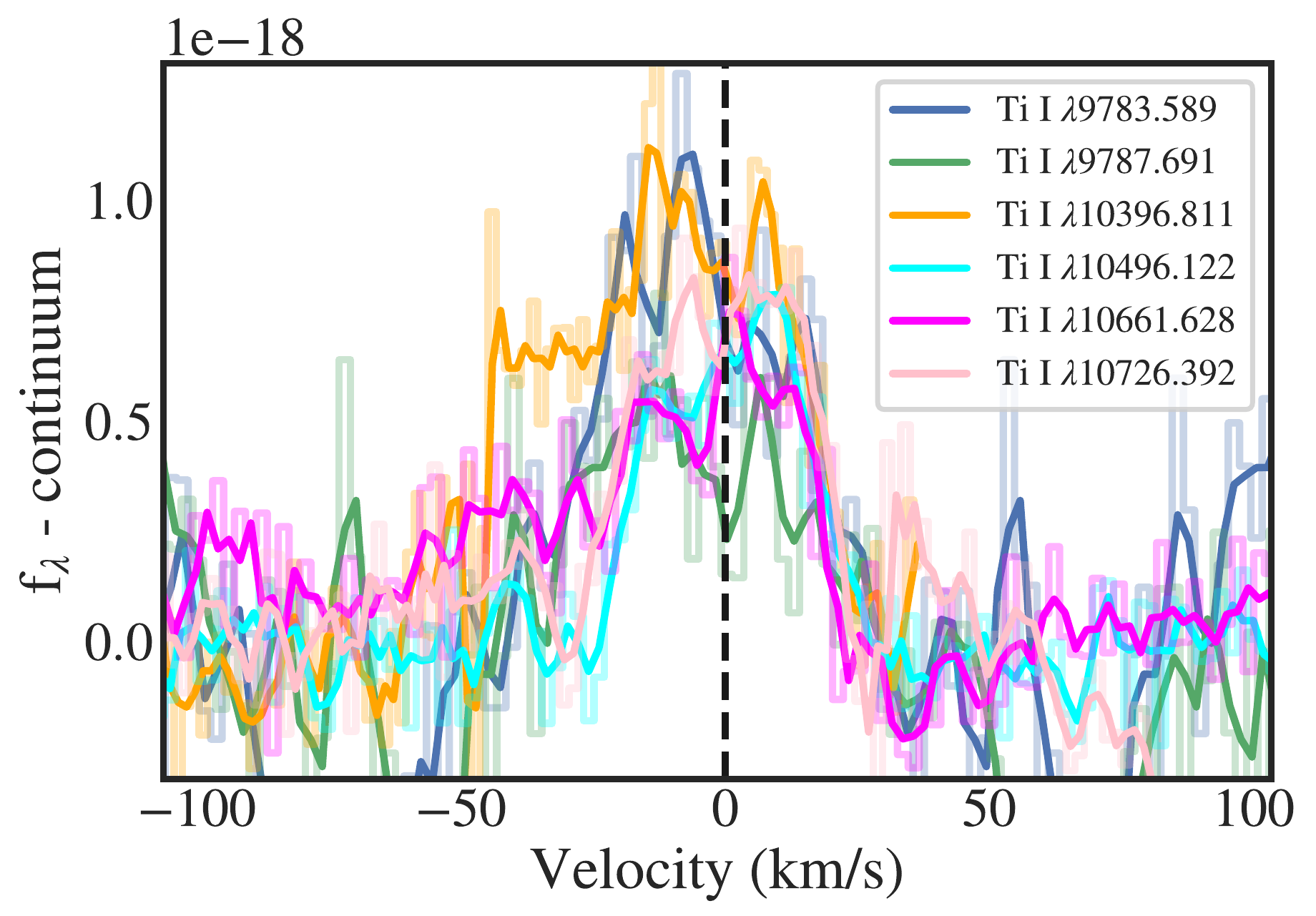}
\caption{\label{HPFFeI} High resolution line profiles of four Fe I lines and six Ti I emission lines on 2020 September 12. These lines are likely originating in the active chromosphere of Gaia 20eae. The fainter step style plots behind the bold lines are the measured spectra. The bold curves are a 3 pixel smoothed spectra shown for clarity.}
\end{figure}

We detect multiple Fe I and Ti I lines in emission from Gaia 20eae during its high accretion phase (Figure \ref{HPFFeI}). These lines typically originate in the active chromosphere, and are relatively symmetric when not blended with other lines. Hence, the peak positions of these lines were used to measure the 20 km s$^{-1}$ radial velocity of Gaia 20eae. The widths of these emission lines are similar across the spectrum.

\begin{table*}
\centering
\caption{Variations of the wind velocities as obtained from blue$-$shifted absorption dips of H$\alpha$, Na I D and H$\beta$ and equivalent width (in ${\AA}$) variations of the optical lines in Gaia 20eae.
The error in the equivalent width is estimated using the relation provided by \citet{1988IAUS..132..345C}.}
\begin{tabular}{@{}c@{ }c|c@{ }c@{ }c|r@{ }r@{ }r@{ }r@{ }r@{ }r@{ }r@{ }r@{}}
\hline
Date & JD &  \multicolumn{3}{c}{Wind velocity (km/s)} &  \multicolumn{7}{c}{Equivalent Width (${\AA}$)}  \\
     &    & H$\alpha$ & Na I D & H$\beta$  & H$\beta$  & Na I D   & H$\alpha$  & [O I]  & \multicolumn{3}{c}{Ca\,{\sc ii}}  \\
     &    &     &  &  & & & &   $\lambda$7773  &$\lambda$8498  & $\lambda$8542  & $\lambda$8662 &  \\
\hline
2020 Aug 29  & 2459091 & $-422$ & $-303$  & $-313$ &$ 4.5\pm0.3$&$ 4.8 \pm   0.6 $&$-5.1  \pm 0.7$&$4.3  \pm 0.6 $&$-5.0  \pm 0.6 $&$-4.8  \pm 0.6$&$-6.1 \pm 0.6  $\\
2020 Aug 30  & 2459092 & $-513$ & $-293$  & $-203$ &$ 7.2\pm0.3$&$ 3.0 \pm   0.3 $&$-4.5  \pm 0.6$&$ -           $&$  -           $&$   -         $&$-             $\\    
2020 Aug 31  & 2459093 & $-512$ & $-362$  & $-200$ &$ 4.1\pm0.3$&$ 2.9 \pm   0.3 $&$-5.6  \pm 0.7$&$ -           $&$  -           $&$   -         $&$-             $\\    
2020 Sept 01 & 2459094 & $-451$ & $-438$  & $-301$ &$ 5.4\pm0.3$&$ 4.2 \pm   0.5 $&$-5.1  \pm 0.7$&$4.5  \pm 0.6 $&$-6.7  \pm 0.8 $&$-7.2  \pm 0.8$&$-6.5 \pm   0.7$\\ 
2020 Sept 07 & 2459100 & $-619$ & $-384$  & $-630$ &$ 8.3\pm1.8$&$ 6.6 \pm   1.2 $&$-8.3  \pm 1.0$&$3.4  \pm 0.6 $&$-7.1  \pm 0.9 $&$-9.3  \pm 1.0$&$-10.3\pm   0.9$\\
2020 Sept 12 & 2549105 & $-$    &  $-$    & $-320$ &$ 4.8\pm0.0$&$ 9.9 \pm   0.0 $&$-8.6  \pm 0.0$&$4.6  \pm 0.0 $&$-6.4  \pm 0.0 $&$-7.8  \pm 0.0$&$-6.6\pm   0.0$\\
2020 Sept 14 & 2459107 & $-516$ & $-356$  & $-407$ &$ 6.3\pm1.1$&$ 8.7 \pm   1.3 $&$-5.6  \pm 0.8$&$4.4  \pm 0.7 $&$-4.6  \pm 0.6 $&$-5.5  \pm 0.7$&$-4.8 \pm   0.6$\\ 
\hline
\end{tabular}
\label{tab:velocity}
\end{table*}

\section{Discussion and Conclusions} \label{pt3}

Gaia 20eae is the farthest discovered FUor/EXor type Class\,{\sc ii} YSO undergoing outburst of $\sim$4.25 mag in $G$ band.
We have found that the present brightening of Gaia 20eae is not due to the dust clearing from our line of sight towards the source but 
due to an intrinsic change in the SED  (warming of the continuum component).
The LC of Gaia 20eae in the quiescent phase is showing a small scale fluctuation
of amplitude of 0.2 mag and  period of $\sim$2 days.

In the outburst phase, Gaia 20eae is showing a transition stage during which most of its brightness ($\sim$3.4 mag) has occurred 
at a short timescale of 34 days with a rise-rate of 3 mag month$^{-1}$.
This rise-rate of Gaia 20eae during the transition stage is greater than most of the
recorded rise rates of FUor/EXor family of sources during the outburst phase, e.g., V899 Mon \citep[0.04-0.15 mag month$^{-1}$;][]{2015ApJ...815....4N},
Gaia 18dvy  \citep[0.42 mag month$^{-1}$ in the Gaia $G$ band;][a newly discovered FUor]{2020ApJ...899..130S} and
V1118 Ori  \citep[1.05 mag month$^{-1}$;][an EXor source]{2017ApJ...839..112G}.
Such difference in timescales of the rise-rates, possibly, implies a different trigger mechanism 
in Gaia 20eae resulting in the present luminosity outburst.   
Once it reached maximum brightness, it slowly started to decay from its maximum brightness with a decay rate of 0.3 mag month$^{-1}$.
The present decay rate is similar to that of bonafide EXor class of sources, EX Lupi and VY Tau, which
returned to its quiescent stages in 1.5-2 years after their maximum brightness state \citep{1977ApJ...217..693H}. The present decay rate is
also similar to that of V899 Mon, which transitioned to a short quiescent state from its 2010 outburst state \citep{2015ApJ...815....4N}.   
The decay rate of Gaia 20eae is surprisingly similar to that of V1118 Ori being 
equal to 0.3 mag/month thus pointing to the fact that a similar relaxing phenomenon is occurring in Gaia 20eae also.
During the outburst phase, Gaia 18dvy is also showing small scale fluctuations having amplitude of $\sim$0.2 mag in all the bands.
Such a short scale accretion variability has also been reported by \citet{2015ApJ...815....4N} for V899 Mon.
Similar fluctuations were observed in FU Ori, and may be
due to flickering or inhomogeneities in the accretion disk
\citep{2000ApJ...531.1028K,2013MNRAS.432..194S, 2020ApJ...899..130S}.

Few interesting spectral features  of Gaia 20eae are tabulated in Table \ref{tab:gaia20compare} to classify Gaia 20eae by comparing its property with
different classes of the episodically accretion low mass young stars \citep{1998apsf.book.....H, 2018ApJ...861..145C}.
Most of them  are matching with EXor source but the 
H$\beta$ absorption line  and  P Cygni profile of H$\alpha$ line hint towards the FUor source classification. The P Cygni profile in H$\alpha$ originates from the winds generated due to accretion of matter through accretion funnels by the process of  magnetospheric accretion. During our spectroscopic monitoring, the P Cygni profile of H$\alpha$ line showed substantial variations. We have found that the outflow wind velocity for Gaia 20eae
shows a large scatter which may be due to the intrinsic variation of wind velocity during the outburst phase. 
As H$\alpha$ line originates from the innermost hot zone powered by accretion, the EW of the emission component of H$\alpha$ line
is an approximate indicator of the accretion rate. 
Table \ref{tab:velocity} shows a large variation in EW values of
H$\alpha$ line during our monitoring period. 
Similar to this, EW of O I $\lambda$7773$\rm\AA$ also varied upto $\sim75 \%$.
This indicates towards the highly turbulent accretion activities going on in Gaia 20eae during its outburst phase. 
This is also evident from the short scale fluctuations of the photometric magnitudes observed during the same period.
These properties of Gaia 20eae are similar to that of the V899 Mon which also showed
heavy outflow activities and increase in disc turbulence as it transitioned to its quiescent state after its first outburst \citep{2015ApJ...815....4N}.
Therefore, similar to V899 Mon, the present outburst of Gaia 20eae might be triggered by the magnetic instabilities in magnetospheric accretion.
Gaia 20eae also clearly shows the decaying phase less than 15 months as well as CO band-heads in the emission.
These features suggest that Gaia 20eae broadly resembles the EXor family of sources.

Our high resolution spectrum shows a very strong blue-shifted He $\lambda$10830$\rm\AA$ absorption signature, which indicates a very strong outflow activity in Gaia 20eae.
We have also  detected a red-shifted absorption component in all the  Ca II IR triplet lines, which could be due to the hot in-falling gas in the magnetospheric accretion funnel. As far as we know, this is the first reported direct detection of an infall signature in Ca II IR triplet lines in FUors/Exors family of objects. We believe that is strong evidence for the magnetospheric funnel origin of Ca II IR triplet lines in heavily accreting YSOs.  
Based on this, we have also constrained the viewing angle to be such that it is through the accretion funnel to the footprint on the stellar surface.

\begin{table}
\centering
\caption{Features of Gaia 20eae compared to bonafide FUors and EXors.}
\label{tab:gaia20compare}
\begin{tabular}{p{1.2in}p{.55in}p{.55in}p{.55in}}
\hline
Feature & FUor & EXor & Gaia 20eae \\
\hline
Outburst amplitude (mag) & 4-6 &  2-4 & 4.6\\
Age      &  Class\,{\sc i/ii}    & Class\,{\sc ii}   & Class\,{\sc ii} \\
Luminosity (L$_{\odot}$)& 100-300 & 0.5-20 & 5.9 \\
Reflection Nebulae & Yes & Sometimes & No\\
Decay to quiescence & 20-100 yr & 0.5-2 yr & 1.3 yr\\
H$\beta$ & Absorption & Emission & Absorption \\
H$\alpha$ & P Cygni & Emission & P Cygni  \\
CO(2-0) and CO(3-1)  & Absorption & Emission & Emission\\
\hline
\end{tabular}
\end{table}

\section*{Acknowledgments}

We thank the anonymous reviewer for valuable comments which greatly improved the scientific content of the paper.
We thank the staff at the 1.3m DFOT and 3.6m DOT, Devasthal (ARIES), for their co-operation during observations.
It is a pleasure to thank the members of 3.6m DOT team and IR astronomy group at TIFR for their support
during TANSPEC observations. TIFR$-$ARIES Near Infrared Spectrometer (TANSPEC) was built in collaboration with TIFR,
ARIES and MKIR, Hawaii for the DOT.
We thank the staff of IAO, Hanle and CREST, Hosakote, that made these observations
possible. The facilities at IAO and CREST are operated by the Indian
Institute of Astrophysics, 
This work has made use of data from the European Space Agency (ESA) mission
{\it Gaia} (\url{https://www.cosmos.esa.int/gaia}), processed by the {\it Gaia}
Data Processing and Analysis Consortium (DPAC,
\url{https://www.cosmos.esa.int/web/gaia/dpac/consortium}). Funding for the DPAC
has been provided by national institutions, in particular the institutions
participating in the {\it Gaia} Multilateral Agreement.  The Center for Exoplanets and Habitable Worlds is supported by the Pennsylvania State University, the Eberly College of Science, and the Pennsylvania Space Grant Consortium.
These results are based on observations obtained with the Habitable-zone Planet Finder Spectrograph on the HET.  We acknowledge support from NSF grants
AST-1006676, AST-1126413, AST-1310885, AST-1310875, ATI 2009889, ATI-2009982, AST-2108512  in the pursuit of precision radial velocities in the NIR.  We acknowledge support from the Heising-Simons Foundation via grant 2017-0494.  The Hobby-Eberly Telescope is a joint project of the University of Texas at Austin, the Pennsylvania State University, Ludwig-Maximilians-Universität München, and Georg-August Universität Gottingen. The HET is named in honor of its principal benefactors, William P. Hobby and Robert E. Eberly. The HET collaboration acknowledges the support and resources from the Texas Advanced Computing Center. We thank the Resident astronomers and Telescope Operators at the HET for the skillful execution of our observations of our observations with HPF. 
CIC acknowledges support by NASA Headquarters under the NASA Earth and Space Science Fellowship Program through grant 80NSSC18K1114.
{SS, NP, RY  acknowledge  the support of the Department of Science  and Technology,  Government of India, under project No. DST/INT/Thai/P-15/2019.
DKO acknowledges the support of the Department of Atomic Energy, Government of India, under Project Identification No. RTI 4002. 
}


\vspace{5mm}
\facilities{HCT (HFOSC), DFOT, DOT (TANSPEC, ADFOSC), ARCSAT, HET (HPF, LRS2)}

 \software{astropy \citep{2013A&A...558A..33A},  IRAF \citep{1986SPIE..627..733T,1993ASPC...52..173T},
 DAOPHOT-II~software \citep{1987PASP...99..191S}}



%\appendix

%\section{Appendix information}

%%.epslatex sample63.tex
%%.epslatex sample63.tex
%%.epslatex sample63.tex

\bibliography{gaia}{}
\bibliographystyle{aasjournal}



\end{document}